\PassOptionsToPackage{pdfpagelabels=false}{hyperref} 
\documentclass[fleqn,usenatbib]{mnras}

\usepackage{newtxtext,newtxmath}

\usepackage[T1]{fontenc}
\usepackage{ae,aecompl}

\usepackage[usenames, dvipsnames]{color}
\usepackage{graphicx}	
\usepackage{amsmath}	
\usepackage{multirow} 

\title[{\em Gaia}  SPSS absolute photometry]{The {\em Gaia} spectrophotometric standard stars survey - IV. Results of the absolute photometry campaign. }

\author[G. Altavilla et al.]{G.~Altavilla,$^{1,2}$\thanks{E-mail: giuseppe.altavilla@ssdc.asi.it, giuseppe.altavilla@inaf.it}
S.~Marinoni$^{1,2}$,
E.~Pancino$^{3,2}$,
S.~Galleti$^{4}$,
M.~Bellazzini$^{4}$,
N.~Sanna$^{3}$,
\newauthor
M.~Rainer$^{3}$,
G.~Tessicini$^{4}$,
J.M.~Carrasco$^{5}$,
A.~Bragaglia$^{4}$,
W.J.~Schuster$^{6}$,
G.~Cocozza$^{4}$,
\newauthor
M.~Gebran$^{7}$,
H.~Voss$^{5}$,
L.~Federici$^{4}$,
E.~Masana$^{5}$,
C.~Jordi$^{5}$,
M.~Mongui\'o$^{5}$,
A.~Castro$^{8,9}$,
\newauthor
M.A.~Pe\~{n}a-Guerrero$^{10}$,
A.~P\'{e}rez-Villegas$^{11}$$^{,6}$\\
$^{1}$ INAF -- Osservatorio Astronomico di Roma, Via Frascati 33, 00078, Monte Porzio Catone (Roma), Italy \\
$^{2}$ Space Science Data Center -- ASI, Via del Politecnico SNC, 00133 Roma, Italy\\
$^{3}$ INAF -- Osservatorio Astrofisico di Arcetri, Largo E. Fermi 5, 50125, Florence, Italy\\
$^{4}$ INAF -- Osservatorio di Astrofisica e Scienza dello Spazio, via Gobetti 93/3, 40129 Bologna, Italy\\
$^{5}$ Departament de F\'{i}sica Qu\`{a}ntica i Astrof\'{i}sica, 
Institut de Ci\`encies del Cosmos (ICCUB),  Universitat de Barcelona (IEEC-UB),
                 Mart\'\i\  i Franqu\`es, 1,\\
\,\,                08028 Barcelona, Spain\\         
$^{6}$ Instituto de Astronom\'ia, Universidad Nacional Aut\'onoma de M\'exico, Apartado Postal 106, C.~P.~22800, Ensenada, B.~C., M\'exico \\            
$^{7}$ Department of Physics and Astronomy, Notre Dame University-Louaize, P.O. Box 72, Zouk Mikael, Lebanon\\
$^{8}$  Observatorio Astron\'omico Nacional, Universidad Nacional Aut\'onoma de M\'exico, Apartado Postal 877, C.~P.~22800 Ensenada, B.~C., M\'exico\\
$^{9}$ Consorcio de Investigaci\'on del Golfo de M\'exico, CICESE, Carretera Ensenada-Tijuana No. 3918, Zona Playitas, 22860 Ensenada, B.~C., M\'exico \\
$^{10}$ Space Telescope Science Institute, 3700 San Martin Drive, Baltimore, MD 21218, USA \\
$^{11}$  Universidade de S\~ao Paulo, IAG, Rua do Mat\~ao 1226, Cidade Universit\'aria, S\~ao Paulo 05508-900, Brazil 
}
\date{Accepted XXX. Received YYY; in original form ZZZ}

\pubyear{2020}

\begin{document}
\label{firstpage}
\pagerange{\pageref{firstpage}--\pageref{lastpage}}
\maketitle


\begin{abstract}
We present Johnson-Kron-Cousins $BVRI$ photometry of 228 candidate spectrophotometric standard stars for the external (absolute) flux calibration of {\em Gaia} data.  The data were gathered as part of a ten-year observing campaign  with the goal of building the external grid of flux standards for {\em Gaia} and we obtained absolute photometry, relative photometry for constancy monitoring, and spectrophotometry. Preliminary releases of the flux tables were used to calibrate the first two {\em Gaia} releases. This paper focuses on the imaging frames observed in good sky conditions (about 9100). The photometry will be used to validate the ground-based flux tables of the {\em Gaia} spectrophotometric standard stars and to correct the spectra obtained in non-perfectly photometric observing conditions for small zeropoint variations. The absolute photometry presented here is tied to the Landolt standard stars system to $\simeq$1 per cent or better, depending on the photometric band. Extensive comparisons with various literature sources show an overall $\simeq$1 per cent agreement, which appears to be the current limit in the accuracy of flux calibrations across various  samples and techniques in the literature. The {\em Gaia} photometric {\em precision} is  presently of the order of 0.1 per cent or better, thus various ideas for the improvement of photometric calibration {\em accuracy} are discussed.
\end{abstract}





\section{Introduction}
\label{sec:intro}

\begin{table*}
\caption{Facilities used in the photometric observations of the {\it Gaia} SPSS. The first three columns report the telescope, instrument, and site of each facility used. The last four columns report, respectively: the total number of frames dedicated to imaging, the percentage contributed by each facility, the total number of frames obtained in photometric and good nights, and the percentage contributed by each facility.}\label{tab:tels}
\begin{tabular}{|l|l|l|r|r|r|r|}
\hline
Telescope     &  Instrument   & Location                        & Imaging [\#] & Imaging [\%] & Photometry [\#] & Photometry [\%]  \\
\hline                                                          
ESO NTT 3.6 m &   EFOSC2      & La Silla, Chile                 & 2943 & 10.6 & 1673 & 18.4 \\
TNG 3.6 m     &   DOLoRes     & Roque de los Muchachos,  Spain  & 5337 & 19.2 & 3217 & 35.4 \\
NOT 2.6 m     &   ALFOSC      & Roque de los Muchachos,  Spain  & 1140 &  4.1 &  497 &  5.5 \\
CAHA 2.2 m     &   CAFOS       & Calar Alto, Spain    &  972 &  3.5 &  198 &  2.2 \\
Cassini 1.5 m  &   BFOSC       & Loiano, Italy                   & 6889 & 24.9 &   69 &  0.8 \\
SPM 1.5 m      &   La Ruca     & San Pedro M\'{a}rtir, Mexico    & 8652 & 31.2 & 3440 & 37.8 \\
REM 0.6 m      &  ROSS/ROSS2   & La Silla, Chile                 & 1695 &  6.1 &    0 &  0.0 \\
TJO 0.8 m      &  MEIA         &  Montsec, Spain &   60 &  0.2 &    0 &  0.0 \\ 
\hline
\end{tabular}
\end{table*}

{\em Gaia}\footnote{\url{https://www.cosmos.esa.int/web/gaia/home}} \citep{2016gaiacollaboration} is a cornerstone mission of the European Space Agency (ESA) that has been performing an all-sky survey since the beginning of the scientific operations in July 2014, about 7 months after the lift-off on 2013 December 19. {\em  Gaia} is building the most complete and precise 6-D map of our Galaxy by observing almost 1.7 billion sources from its privileged  position  at the Lagrangian point L2  and its data are already producing major advancements in all branches of astronomy, from the solar system and stellar structure to the dynamics of the Milky Way and cosmology  \citep{PERRYMAN2001,MIGNARD2005,pancino20}. A description of the satellite and a list, inevitably non-exhaustive, of the mission scientific goals  is available in \citet{2016gaiacollaboration}, while an up-to-date  list of refereed {\em Gaia} papers since launch (more than 3600 as of July 2020) is maintained on the ESA webpages\footnote{\url{https://www.cosmos.esa.int/web/gaia/peer-reviewed-journals}}.

The very first products provided to the astronomical community were the photometric science alerts\footnote{\url{https://gsaweb.ast.cam.ac.uk/alerts}}, with the first transient announced on 2014 August 30, whereas solar system alerts for new asteroids\footnote{\url{https://gaiafunsso.imcce.fr/}} started in 2016. The first intermediate data release ({\em Gaia} DR1) was published on 2016 September 14 \citep{2016GDR1} and the second intermediate data release  ({\em Gaia} DR2) took place on 2018 April 25 \citep{2018GDR2}. The next releases will give increasingly accurate and complete sets of astrophysical data. The upcoming {\em Gaia} third and following releases are announced by ESA in their regularly updated `data release scenario'\footnote{\url{https://www.cosmos.esa.int/web/gaia/release}.}. The nominal five-year mission ended on 2019 July 16, but the {\em Gaia} micro-propulsion fuel exhaustion is foreseen only by the end of 2024. A two-years mission extension was approved and further extensions are under discussion. The final {\em Gaia} data release  will presumably be based on 8, hopefully 10, years of observations, providing a new astrometric foundation for astronomy. 

{\em  Gaia} performs astrometric, photometric and spectroscopic measurements by means of two telescopes, observing in directions separated by 106.5$^{\circ}$ (the ``basic angle''), but sharing the same focal plane, one of the largest ever sent to space, that hosts five instruments: the Sky Mapper, the Astrometric Field, the Blue Photometer, the Red Photometer and the Radial Velocity Spectrometer \citep{2016gaiacollaboration}. {\em  Gaia} spectrophotometry in particular comes from different instruments. The first is the Astrometric Field, an array of 62 CCDs operating in white light and defining the {\em Gaia} $G$~band, determined by the reflectivity of the mirrors and  the quantum efficiency of the detectors. The $G$ band covers the optical range (330 -- 1050~nm), peaking around 600~nm. The colour information comes instead from the Blue and Red Photometers (BP and RP), providing low resolution dispersed images in the blue (330 -- 680~nm, $30 \lesssim R \lesssim 100$) and in the red (640 -- 1050~nm, $60 \lesssim R \lesssim 90$) domains. In particular, extremely precise three-band photometry was published in {\em Gaia} DR2 \citep{Evans18}, with sub-millimagnitude internal uncertainties. The photometry was obtained by integrating data from the astrometric field and the spectrophotometers, producing the $G$, $G_{\rm{BP}}$, and $G_{\rm{RP}}$ integrated magnitudes. The quality of the data is so exquisite and unprecedented that subtle systematic effects at the mmag level cannot be tested with existing data-sets, because no photometric survey has such similar precision \citep{Evans18}. 
The upcoming {\em Gaia} release will be split into two releases: an early release (EDR3), publishing improved integrated magnitudes,  and a DR3 release that will include also the epoch-averaged BP and RP spectra for a selected subset of the observed sources.

As outlined by \cite{PANCINO2012} and \cite{CARRASCO2016}, while the {\em Gaia} internal calibration is based solely on {\em Gaia} data, external spectrophotometric standard stars (hereafter SPSS) are required to report the calibration of {\em Gaia} spectra and integrated photometry onto a physical flux scale. Given the instrument complexity, a large set of SPSS (about 200) is required, adequately covering a wide range of spectral types. The best existing set in the literature in 2006 was CALSPEC \citep{bohlin14}, but it only contained about 60 stars and the spectral type coverage was limited for our purpose. The CALSPEC set has evolved significantly over time, most notably by refining the quality of the  spectra, their calibration, and by adding three very red objects \citep{bohlin19}, but still  contains only about 90 spectra covering the entire {\em Gaia} wavelength range.  Moreover, even the latest CALSPEC set lacks K and M-type stars, which are mandatory for calibrating {\em Gaia}. For this reason in 2006 we started an extensive ground based observing campaign, to provide a suitable SPSS grid, extending the list of the CALSPEC stars that matched our original criteria, both by doubling the number of sources, and by better sampling the colour space towards intermediate and red colours, especially in the FGK stars range and in the early M-type stars. Our spectrophotometric calibration is tied to the CALSPEC scale by using their three hydrogen white dwarfs (G191-B2B, GD71,  GD153, see \citealt{BOHLIN1995}) as calibrators of the {\em Gaia} SPSS grid.

In this paper we present the results of our absolute photometry campaign. Given the difficulty of obtaining a large amount of very high quality spectrophotometry, we decided to gather photometric observations of our SPSS candidates\footnote{ The final selection of SPSS for the calibration of {\it Gaia} is done on the flux tables obtained from the spectrophotometry, therefore in this paper all SPSS will be considered candidates. } with the following goals: 
\begin{enumerate} 
\item{to refine the literature photometry of the SPSS candidates, that often had large uncertainties;} 
\item{to validate the flux tables obtained from the spectrophotometry under photometric conditions \citep{NS-001}  \footnote{  The Gaia Technical Notes cited in this work are publicly available on the ESA webpage: \url{https://www.cosmos.esa.int/web/gaia/public-dpac-documents}. };}
\item{to validate -- and when necessary to adjust -- the flux calibration of the spectra obtained under reasonably good but not perfectly photometric conditions.} \end{enumerate}

This is the fourth paper of the {\em Gaia} SPSS series. The first paper presented the ground-based project, the candidate SSPS, and some preliminary results \citep{PANCINO2012}; the second paper presented a detailed study of instrumental effects and the techniques used to minimize their impact \citep{ALTAVILLA2015}; the third paper presented the results of our constancy monitoring campaign (within $\pm$5~mmag), including all variability curves, and led to the exclusion of several candidate SPSS
\citep[some were widely used standards,][]{MARINONI2016}. In this paper, we focus on the imaging data obtained during photometric nights, and we present the absolutely calibrated magnitudes of 228  stars, including  198 candidates SPSS and 30 rejected candidate SPSS. The final release of flux tables based on the spectrophotometric observations will be presented in a forthcoming paper.

The paper is organized as follows: in Section~\ref{sec:obs} we present the observing campaign and the data analysis; in Section~\ref{sec:photcal} we present the photometric calibration; in Section~\ref{sec:valid} we validate our results, comparing them with several literature data collections; in Section~\ref{sec:disc} we discuss  the present status of spectrophotometric calibrations in the literature and future prospects, and we finally draw our conclusions.

\section{Observations and data reduction} 
\label{sec:obs}

The observing campaign for the {\em Gaia} SPSS
\citep[see also][]{PANCINO2012} started in the second half of 2006 with 5 pilot runs, used to refine the observing strategy. The main campaign, for both the spectroscopic and photometric data acquisition, started in January 2007 and ended in July 2015, for a total of 967 different observing nights at 8 different telescopes (see Table~\ref{tab:tels}). 
The data collected during the whole campaign amount to $\simeq$70600 scientific frames ($\simeq$64000 images and $\simeq$6600 spectra).
Here we focus on the imaging data gathered in good observing conditions (about 9100 frames). 

\subsection{Photometric observations and data summary}

From the whole body of imaging observations, we extracted more than 27000 individual photometric catalogues, one for each observed imaging frame with sufficient quality ($\simeq$5000 of which with   \citealt{LANDOLT1992a} standard stars fields measurements, $\simeq$22000 related to candidate SPSS). The breakdown of the photometric catalogues for  each  facility used is shown in Table \ref{tab:tels}, as well as the number of catalogues that were specifically used in this paper (about 9000).

We obtained reliable  absolutely calibrated magnitude measurements for 228 candidate SPSS  in the Johnson-Kron-Cousins  $BVR$ bands and in some cases in the  $I$ band as well. The data used here were gathered in 95 observing nights under clear sky conditions. Of these, three nights did not have Landolt fields observations of sufficient quality to determine a reliable night solution, 38 were judged non-photometric, 22 were reasonably good, and 32 were judged photometric (see Section~\ref{sec:qc} for more details). We observed $\simeq 88$ per cent of the  SPSS candidates at least once under photometric conditions while $\simeq 9$ per cent were observed under good but not fully photometric conditions. Six additional stars observed only in non-photometric conditions are not considered here. The relative photometry data taken for time-series monitoring of the SPSS constancy were presented by \cite{MARINONI2016} and will not be discussed here.

\begin{table*}
\caption{Atmospheric extinction coefficients obtained in our photometry campaign. The number of nights used for the statistics is also indicated.}\label{tab:km}
\begin{tabular}{|l|l|l|l|l||l|l|l|l|l|l}
\hline
Site & Telescope & \# nights &  $k_B$ &  $\delta k_B$  &  $k_V$ &  $\delta k_V$  & $k_R$ &  $\delta k_R$  &  $k_I$ &  $\delta k_I$ \\
  & & & \multicolumn{2}{c|}{ (mag/airmass) } & \multicolumn{2}{c|}{(mag/airmass)} & \multicolumn{2}{c|}{(mag/airmass)} & \multicolumn{2}{c|}{(mag/airmass)} \\
\hline
Calar Alto & CAHA 2.2 m & 3 & 0.194 & 0.023 & 0.116 & 0.017 & 0.067 & 0.032 &  - &  -\\ 
Loiano & Cassini 1.5 m & 2 & 0.229 & 0.014 & 0.143 & 0.006 & 0.109 & 0.006 &  - &  -\\
Roque & NOT 2.6 m & 2 & 0.224 & 0.021 & 0.134 & 0.013 & 0.073 & 0.026 &  - &  -\\ 
La Silla & NTT 3.6 m & 10 & 0.230 & 0.015 & 0.117 & 0.013 & 0.076 & 0.009 &  - &  -\\ 
San Pedro M\'artir & SPM 1.5 m & 25 & 0.213 & 0.021 & 0.133 & 0.043 & 0.087 & 0.034 & 0.037 & 0.003\\
Roque & TNG 3.6 m & 12 & 0.227 & 0.034 & 0.143 & 0.024 & 0.101 & 0.016 &  - &  -\\
\hline
\end{tabular}
\end{table*}

The lists of candidate SPSS and rejected candidates presented here  (see Tables~\ref{tab:phot},~\ref{tab:rej}) does not correspond entirely to those by \citet{PANCINO2012}. On the one hand, a few more candidate SPSS were rejected since 2012 because of variability, either detected by \citet{MARINONI2016} or by other literature sources, or because of other problems such as close companions, problems during observations, and so on (see notes to Table~\ref{tab:rej}). In particular, 14 candidate SPSS  originally included in \citet{PANCINO2012} were rejected, while two new candidates were observed after 2012 and included only here. On the other hand, not all good SPSS candidates\footnote{At this point, we consider as good SPSS candidates those matching the initial criteria defined by \citet{PANCINO2012}, that were not rejected so far based on new information, and that have good spectroscopic observations.} were observed under photometric conditions, and thus some do not appear in this paper. In particular, we present here absolute photometry for 228 SPSS candidates, of which 30  are now rejected and presented separately\footnote{One of the good SPSS candidates, SPSS\,355 or SDSS\,J125716+220059, has good photometry, but no spectroscopic observations, so it is unlikely to be included in the final {\em Gaia} SPSS grid.}. Finally, for two good SPSS candidates, SPSS\,159 (WD\,0050-332) and SPSS\,219 (WD\,0106-358), we could not obtain reliable absolute photometry, unfortunately.

\begin{figure} 
\includegraphics[clip,width=\columnwidth]{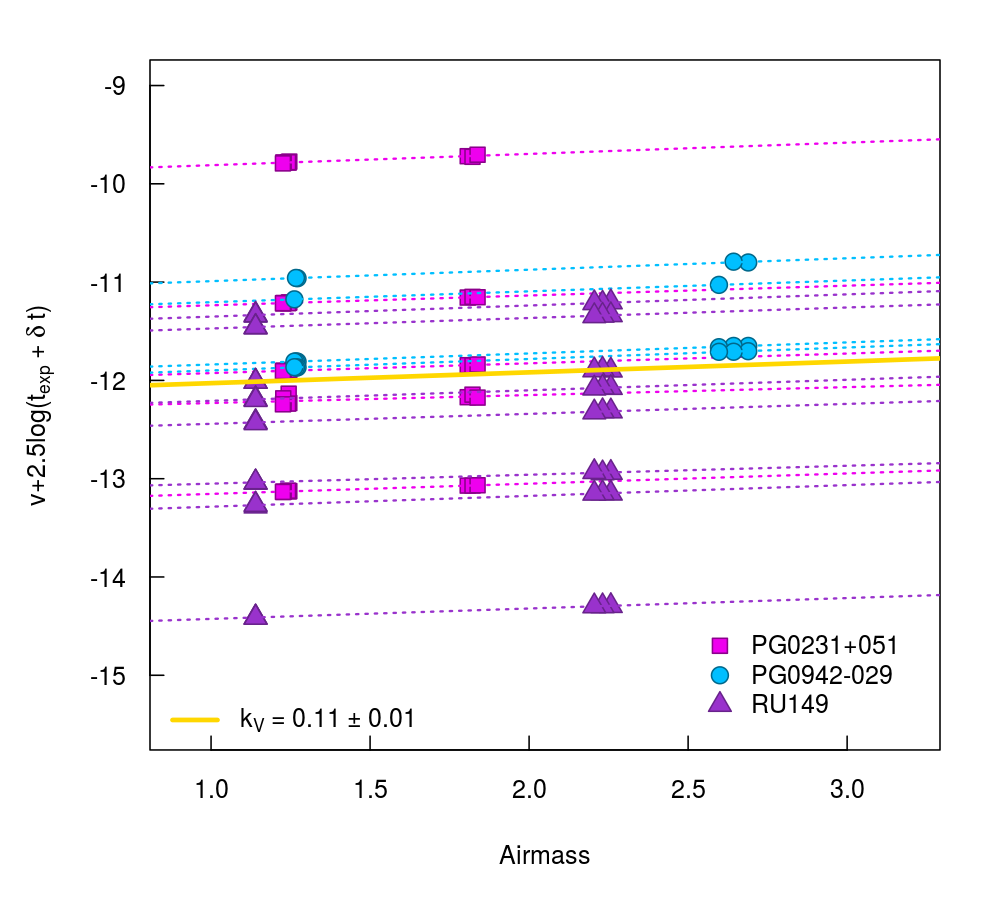}
\vspace{-0.3cm}
\caption{Example of determination of the extinction coefficient $k_{v}$ in the $V$ band, for the night of 2008 November 26, at the ESO NTT telescope. Dotted lines -- for each star in a given Landolt field -- are linear fits of the stars's instrumental $v$ magnitude corrected for exposure time, as a function of effective airmass. Each slope represents an estimate of $k_{v}$: their median and semi-interquartile range provide a robust estimate of $k_{v}$ for the night (yellow thick line). }
\label{fig:am}   
\end{figure}

\subsection{Data reduction and aperture photometry}
\label{sec:datared}

To allow for the most homogeneous and accurate reduction possible, a specific characterization of all the instruments used was initially carried out \citep{ALTAVILLA2015}, including a detailed study of the stability  of the  calibration frames (i.e. master bias, dark, master flat and bad pixel mask). The imaging frames used here were thus pre-reduced, by removing instrument signatures, following the procedure described in detail by \citet{SMR-001,MARINONI2016}. Briefly, the procedure includes the standard bias (and overscan, when available) subtraction and flat field normalization. Dark current correction was not needed for the six facilities used for absolute photometry. Bad pixels were flagged and used in an accurate quality control procedure \citep[hereafter QC, see also][for details]{SMR-003} of the master frames  considering saturation, signal-to-noise ratio (SNR), defects, etc. Our procedures were mostly based on \textsc{iraf}\footnote{\textsc{iraf} was distributed by the National Optical Astronomy Observatories, which are operated by the Association of Universities for Research in Astronomy, Inc., under cooperative agreement with the National Science Foundation.} \citep{tody86,tody93}, but also on in-house scripts.

Once the data were corrected for instrumental signatures, we proceeded to measure aperture photometry on all imaging frames taken on nights where also standard stars from the \citet{LANDOLT1992a} catalogues were observed. We used \textsc{SExtractor} \citep{BERTIN1996} to sum up the flux contained in large apertures, chosen as six times the full width at half maximum (FWHM) of the stellar profiles, to minimize flux losses. The same criterion was applied to SPSS and Landolt calibration observations. The procedure is described in detail in \cite{SMR-003}. We profited, among other things, from the variety of output flags from \textsc{SExtractor} to perform additional QC checks as detailed by \citet{SMR-003}, related to both the frame and the stars, concerning saturation, SNR, seeing and focus, bad pixels, the number of good Landolt stars in standard frames, and so on.

The resulting photometric catalogues, one for each frame, contain instrumental magnitudes for each relevant star in each frame, either Landolt standards or SPSS candidates, depending on the type of observation. 
The instrumental magnitudes are defined as \begin{equation}  m = -2.5~\log_{10} DN_{\rm{star}}\label{eq:instm}\end{equation}
\noindent where $m$ is the instrumental magnitude (in our case $b$, $v$, $r$, or $i$) and $DN_{\rm{star}}$ is the  signal of the star, here represented by the total CCD counts measured inside the aperture after subtracting the sky background.

\section{Photometric calibration} 
\label{sec:photcal}

To transform the instrumental aperture magnitudes to the Landolt system, we used a standard procedure \citep[see, e.g.,][]{stetson19}: we first computed the night solutions using observations of Landolt standard fields and then applied the solutions to the SPSS instrumental magnitudes to obtain the  absolutely calibrated magnitudes. Before computing the  absolutely calibrated magnitudes, however, we performed a detailed assessment of the night quality, as described in  the following sections.

\subsection{Extinction coefficients}
\label{sec:ext}

\begin{figure}
\includegraphics[clip,width=\columnwidth]{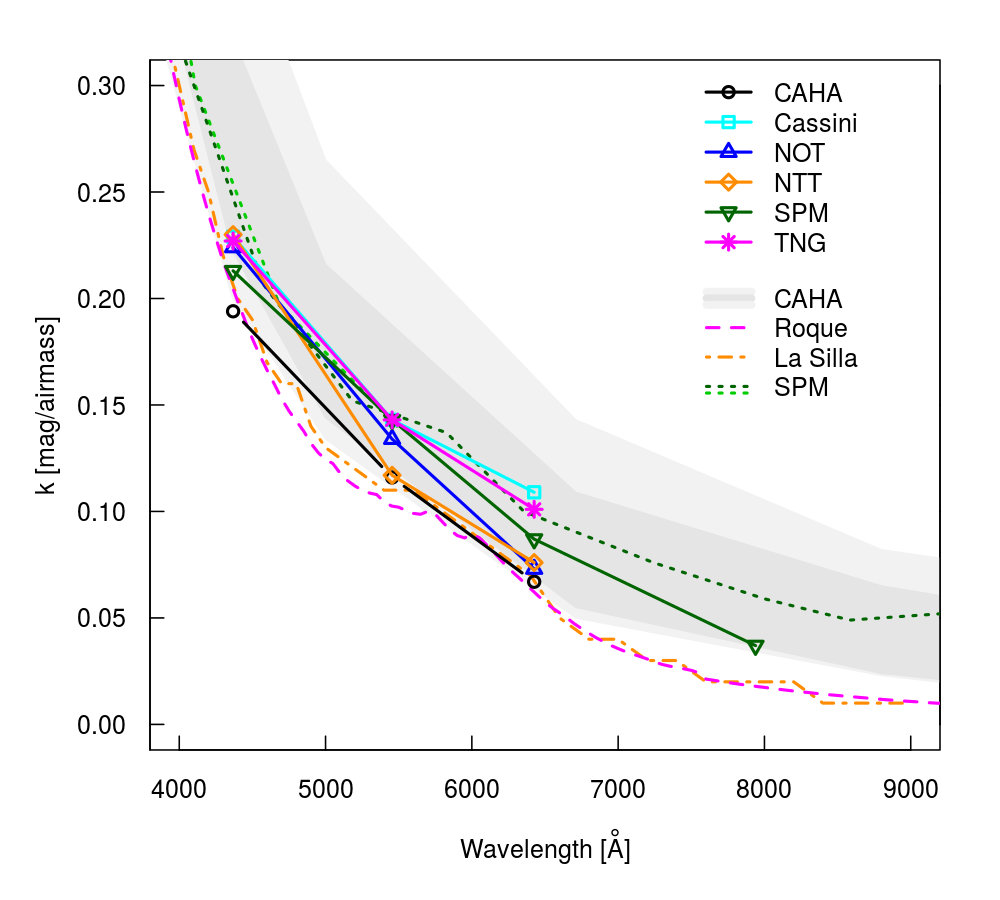}
\vspace{-0.3cm}
\caption{Extinction coefficients $k_M$ plotted as a function of the  effective wavelengths of the  $BVRI$  passbands  by \citet[][]{BESSELL2012}, measured at the six observing sites, as detailed in the upper legend.  Literature data are also shown for comparison (see Section~\ref{sec:ext} for references), as detailed in the lower legend.}
\label{fig:km}   
\end{figure}

The actual calibration process was split into two separate steps. In the first step, we took into account the atmospheric and exposure-time effects using the following equation
\begin{equation} m'=m +2.5 \log_{10} (t_{\rm{exp}} + \delta t) - k_M X\label{eq:am}\end{equation}
\noindent where $ m'$ is the corrected instrumental magnitude (in our case  $ b'$, $ v'$, $ r'$, or $ i'$), $t_{\rm{exp}}$ is the exposure time, $\delta t$ is the shutter effect, necessary to obtain the effective exposure time \citep[determined as in][]{GA-004,ALTAVILLA2015}, $k_M$ is the extinction coefficient in the $M$ band (in our case $k_B$, $k_V$, $k_R$, or $k_I$), and $X$ is the effective airmass\footnote{The effective airmass is computed with the \textsc{iraf} task {\it setairmass} according to the formula described in \cite{STETSON1989} to approximate the mean airmass $\bar{X}$ of long exposures better than the mid-observation value $X_{1/2}$: $\bar{X} = (X_0 + 4 X_{1/2} + X_1)/6 $, where $X_0$ and  $X_1$ are the airmasses at the beginning and at the end of the exposure respectively.}.
To compute the atmospheric extinction coefficient for each band and each observing night, we used repeated observations of the same stars in a few Landolt standard fields, observed at different $X$, to compute  $i$ independent estimates of the extinction coefficient, $k_{M_i}$, as exemplified in Figure~\ref{fig:am}. We then used their median as a robust estimate of $k_M$ for the night in each band, and their inter-quartile range as the related uncertainty $\delta k_M$. 
To be noted that standard star fields were occasionally observed also at large zenith distances to probe the atmospheric absorption  in a wide airmass range  whereas scientific targets were observed preferably as high as possible above the horizon to ensure
high quality measurements.
Having several nights at different observing sites, we could compare our typical extinction coefficients with literature estimates for each observing site.  To compute typical coefficients, we used the average of 2--25 nights, depending on the site, as detailed in Table~\ref{tab:km}. Unfortunately, our data are not sufficient to properly sample seasonal variations, which might be relevant and characterized by lower values in winter \citep{METLOV2004,SANCHEZ2007}. For Calar Alto, we used extinction data by \citet{SANCHEZ2007}, who also provided uncertainties; for La Palma, we found useful data, valid only for very clear nights, without aerosols, in the RGO/La Palma technical note n.~31, and detailed in the William Herschel Telescope website\footnote{\url{http://www.ing.iac.es/astronomy/observing/manuals/html\_manuals/wht\_instr/pfip/prime3\_www.html}}; for typical nights the values could be increased by 0.05  mag/airmass; for La Silla we used values from the ESO 1993 user manual\footnote{\url{https://www.eso.org/sci/observing/tools/Extinction.html}}; for San Pedro M\'artir we used the values by \citet{SCHUSTER2001}. The results of the comparison are summarized in Figure~\ref{fig:km}, displaying excellent agreement.
To be consistent with literature data, in Figures~\ref{fig:km},\ref{fig:NTTZP}  we computed $k_B$ as described above, 
neglecting color dependencies, but for the actual calibration of the $B$-band magnitudes
we  expanded eq.~\ref{eq:am} to account for  secondary extinction correction effects:
\begin{equation} m'=m +2.5 \log_{10} (t_{\rm{exp}} + \delta t) - (k'_M + k''_M c) X  \label{eq:amkb}\end{equation}
where $k'_M$ and $k''_M$ are the primary  and secondary extinction coefficients respectively,
$c$ is the observed color index, that we approximated as the difference between $b'$ and $v'$ computed  
with eq.~\ref{eq:am} (see also Sec.\ref{ZPandCT}).
The second order extinction  coefficients can be determined applying  eq.~\ref{eq:amkb} to a pair of  stars with extreme colors  
observed in the same field  and spanning a wide airmass range, e.g. the reddest and the bluest stars in a  standard star field 
repeatedly observed during the night. Once  $k''_B$ is known, we can apply eq.~\ref{eq:amkb} again, this time to stars spanning a wide airmass range,
regardless of their colour, e.g., stars in a  standard star field  repeatedly observed during the night, to determine $k'_B$,  
as exemplified by \cite{BUCHHEIM2005}.
Due to the difficulties in measuring $k''_B$, we used the mean value obtained over several nights, while for  $k'_B$ we used the values
obtained night by night. The  average values of $k'_B$ and $k''_B$ for each telescope\footnote{The SPM 1.5m adopted different CCDs along the observing campaign but the secondary extinction coefficients computed for each CCD are all consistent within the uncertainties, hence we used a single value.}
are shown in Table~\ref{tab:kBsecondo}. It is to be noted that  the  $k'_B$ values corresponding to $k''_B$ are not the  $k_B$ values reported in Table~\ref{tab:km} (in fact the slope mentioned in Figure~\ref{fig:am} corresponds to the term $k'_M + k''_M c$ in eq.~\ref{eq:amkb}).
According to our data, the $k'_B$ values  are roughly $\sim 0.03$ mag/airmass larger, on average, than $k_B$\footnote{The large $k'_B$ value measured in Loiano is based on a single, possibly problematic, night.}.
The $B$-band second order extinction term makes blue stars dim faster than red stars as they approach the horizon. However, the secondary extinction coefficients in $VRI$ are smaller than in $B$ and smaller than the uncertainties, so we neglected them as normally done \citep{SUNG2013}.

\begin{table*}
\caption{B-band primary and secondary extinction coefficients obtained in our photometry campaign. The number of nights used for the statistics is also indicated.}\label{tab:kBsecondo}
\begin{tabular}{|l|l|l|l|l|l|l|l|}
\hline
Site  & Telescope & \# nights $k'_B$ &
$k'_B$  &  $\delta k'_B$  & \# nights $k''_B$ & 
$k''_B$ &  $\delta k''_B$ \\
 & & &  \multicolumn{2}{c|}{(mag/airmass)} & &\multicolumn{2}{c|}{(mag/airmass)} \\
\hline
Calar Alto         & CAHA 2.2m     & 5  & 0.26   &0.05  & 5  & -0.03  &0.03  \\
Loiano             & Cassini 1.5 m & 1  & 0.41  &0.01 & 1  & -0.01  &0.02\\
Roque              & NOT 2.6 m     & 2  & 0.26   &0.02  & 2  & -0.04  &0.01  \\
La Silla           & NTT 3.6 m     & 11 & 0.25   &0.02  & 11  & -0.03  &0.02 \\
San Pedro Mart\'ir & SPM 1.5m      & 25 & 0.24   &0.02  & 24  & -0.03  &0.02 \\
Roque              & TNG 3.6m      & 12 & 0.24   &0.04  &  8 & -0.03  &0.02  \\
\hline
\end{tabular}
\end{table*}

\begin{figure} 
\includegraphics[clip,width=\columnwidth]{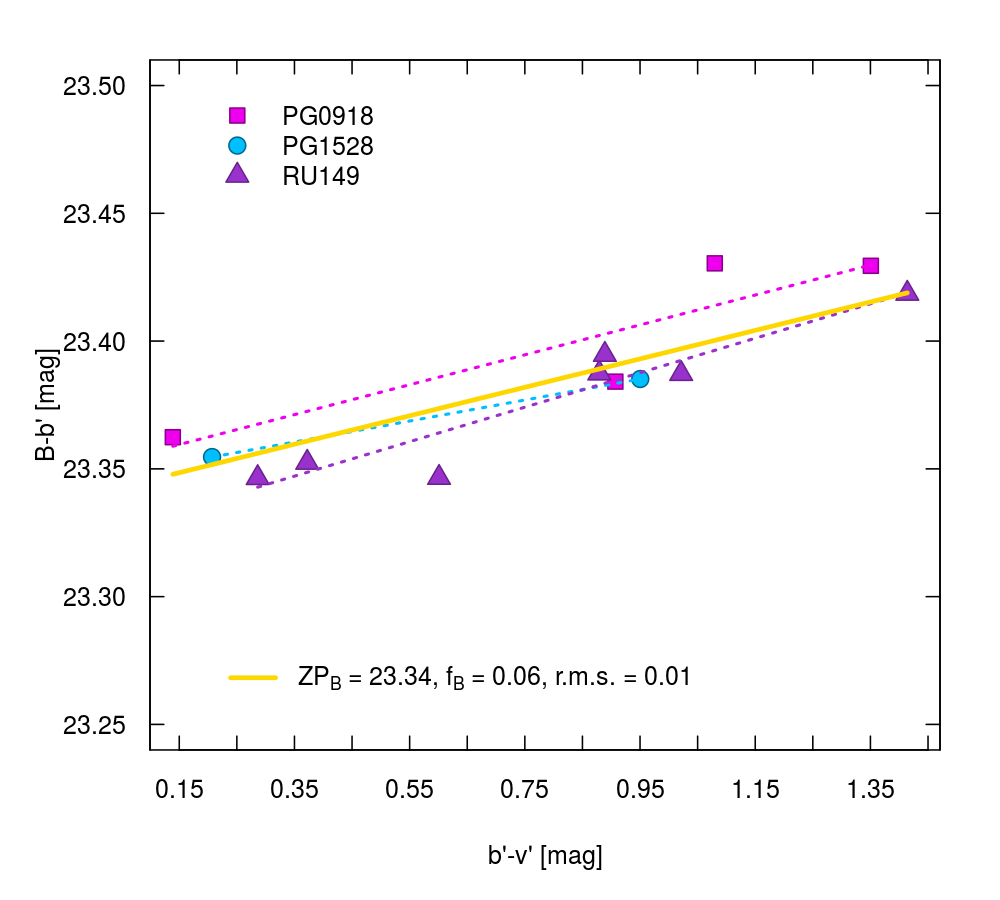}
\vspace{-0.3cm}
\caption{Example of the computation of the colour term $f_B$ and zero point $ZP_B$ for the San Pedro M\'artir observations of 28 January 2009. Each of the three Landolt fields observations are represented as filled symbols of a different colour. The comparison of linear fits for each field separately (coloured dotted lines) allows to verify that they provide compatible estimates. The global $f_B$ and $Z_B$ for the night are indicated at the bottom of the plot, and represented by a solid line.}
\label{fig:sol}   
\end{figure}
\subsection{Zero-points and colour terms}\label{ZPandCT}

Once the instrumental magnitude was corrected for exposure time and extinction, the photometric solution for the night could be computed with the equation 
\begin{equation}
M =  m'  + f_M\,c + ZP_M
\label{eq:sol}
\end{equation}
\noindent where $M$ is the calibrated magnitude (in our case $B$, $V$, $R$, or $I$); $f_M$ is the colour-term of the calibration equation; $c$ is a colour built using the observed instrumental magnitude $m'$ and a nearby  passband (e.g., $b'-v'$ for the $V$-band calibration)\footnote{We use a first-order polynomial approximation of the actual spectral energy distribution in terms of the star's colour, because a few experiments showed that it was not necessary to include a second-order term.} that takes into account the dependence of the solution on the spectral energy distribution of the considered star; and $ZP_M$ is the absolute zero point of the solution. Slight deviations between the standard stars' photometric system and the photometric system actually adopted in the observations, that depends on the passbands and on the overall optical throughput of the telescope and the camera, are expected, as extensively discussed in Sections~\ref{sec:landolt} and \ref{sec:calspec}. Such differences are tracked and corrected  through observations of several standard stars, spanning a wide range of colours.
If the standard stars' photometric system and the photometric system actually adopted in the observations perfectly match, $f_M$ should be ideally equal to zero. Nevetheless slight deviations between the two systems,  that depend on the passbands and on the overall optical throughput of the telescope and the camera actually used, are expected, as extensively discussed in Sections~\ref{sec:landolt} and \ref{sec:calspec}. In this case the colour-term can be computed through observations of several standard stars, spanning a wide range of colours.

Some authors write Equation~\ref{eq:sol} using the standard tabulated colour $C$ \citep[in our case, from][]{LANDOLT1992a} instead of the instrumental colour $c$ (e.g, $B-V$ instead of $b'-v'$). The problem of the choice of the independent variable in Equation~\ref{eq:sol} was discussed at length by \citet{H81}. In practice, to find the solution an ordinary least-square algorithm is used, which requires various conditions, including that the independent variables are error-less  and the dependent ones are homoscedastic. These conditions are never fully satisfied, but are less violated by the Landolt tabulated magnitudes than by the observed ones. Hovever, \citet{isobe} suggest that the reason to adopt tabulated magnitudes as the independent variables is not so compelling, and the use of the instrumental magnitudes as the independent variable can also be a viable method as far as the errors on instrumental magnitudes are comparable to the errors on the tabulated  absolutely calibrated  magnitudes. The typical uncertainties on our instrumental magnitudes are about 1--2 per cent, fully compatible with the Landolt ones (see Section~\ref{sec:landolt}). Therefore, we used the observed instrumental magnitudes in Equation~\ref{eq:sol}   as independent variables.

\begin{figure*} 
 \includegraphics[clip,width=\textwidth]{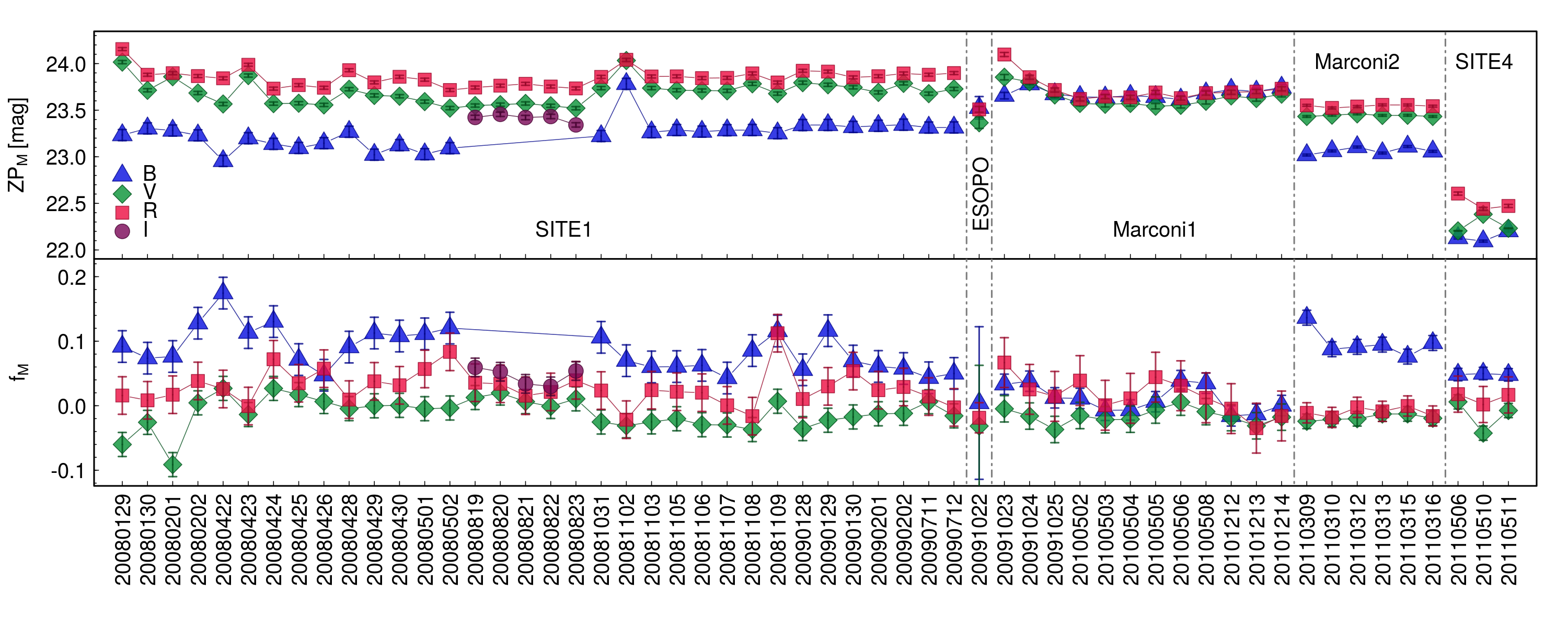}
\caption{Temporal evolution of the zero points $ZP_M$ and colour terms $f_M$ (Equation~\ref{eq:sol}) as measured with the 1.5 m telescope at the San Pedro M\'artir Observatory from January 2008 to May 2011. The names of the five CCDs mounted at the La Ruca imager during this period are shown.}
\label{fig:spmZPCT}   
\end{figure*}

\begin{figure*} 
\includegraphics[clip,width=\textwidth]{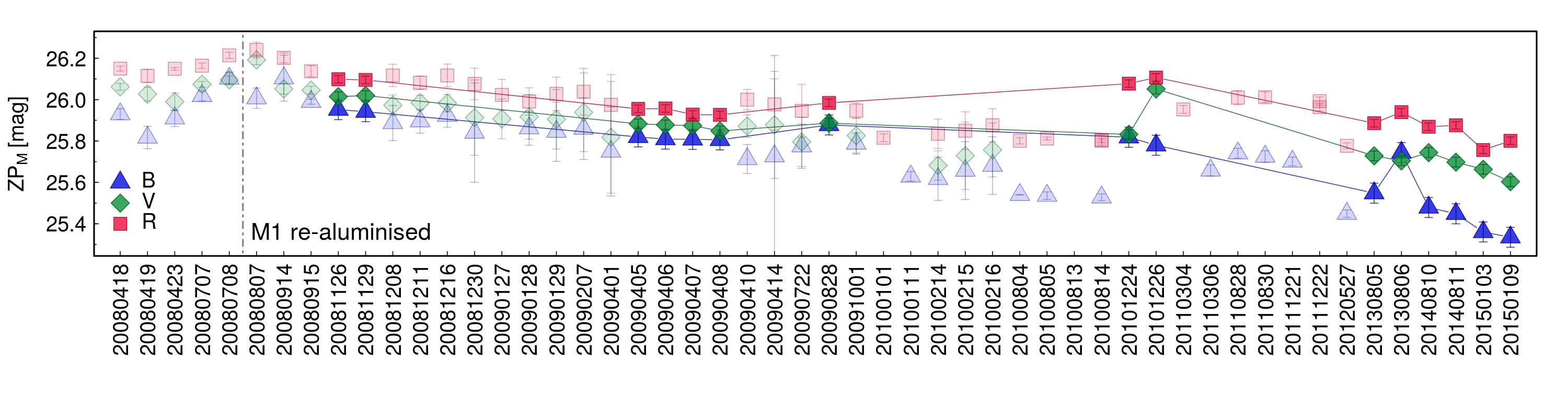}
\caption{Temporal evolution of the zero points $ZP_M$ as measured with the NTT telescope at the ESO La Silla  Observatory from November 2008 to January 2015. Light coloured symbols represent archival ESO data (see text for details), while full coloured symbols show our measurements.}
\label{fig:NTTZP}   
\end{figure*}

In practice, we proceeded as follows: (i) we used Equation~\ref{eq:sol} separately on each Landolt field observed during the night; (ii) we obtained $f_M$ and $ZP_M$ for each photometric band using ordinary least squares; (iii) we verified that the solutions for each field in the same night are compatible with each other (see Figure~\ref{fig:sol}), otherwise it might mean that the sky conditions have changed or that the discrepant fields have problems; (iv) we averaged the $f_M$ and $ZP_M$ obtained for each band and used the root mean square deviation to represent the uncertainties $\delta f_M$ and $\delta ZP_M$.

To illustrate the quality of the results, we show in Figure~\ref{fig:spmZPCT} the trends of the $ZP_M$ and $f_M$ in the four available bands for our observing nights at San Pedro M\'artir. During the years, the instrument was refurbished four times, with a change of detector, and this has caused visible jumps in the typical values in each band. Moreover, random variations due to variable sky conditions are well visible. Figure~\ref{fig:NTTZP} similarly shows the behaviour of $ZP_M$ for our observing runs at the ESO NTT in La Silla. In this case, we could also compare our results with the ones in the ESO Archive\footnote{\url{http://www.ls.eso.org/sci/facilities/lasilla/instruments/efosc/archive/efosczp.dat}}. Small difference in the comparison could be caused by different imaging mode (different gain) and other minor effects. In any case, the agreement between our $ZP_M$ data for NTT and the ESO ones is excellent.

\subsection{Night quality control}
\label{sec:qc}

To evaluate whether a given observing night was useful for our purpose, we performed a dedicated quality assessment. A detailed description of the adopted procedures can be found in \citet{GA-007}; here below we provide a brief summary. 

A complete dataset for the determination of $k_M$ consists of at least three different Landolt standard fields observed in each of three bands (either $BVR$ or $VRI$ in our case), with an airmass difference of $\Delta X \geq 0.2$. A dataset cannot be useful when two out of three bands do not have useful data, i.e., when $k_M$ is only available for one band. All other cases were considered intermediate: the data are not optimal, but still could be useful. Whenever $k_M$ could be computed for at least two bands, we used its uncertainty to decide whether its determination was good ($\delta k_M \leq 0.03$~mag/airmass in all bands), bad ($\delta k_M \geq 0.06$~mag/airmass in all bands) or intermediate (all other cases).

A similar approach was followed to assess whether the available datasets were adequate to compute reliable $ZP_M$ and to assess their quality. Here, instead of $\Delta X$ we used the time coverage $\Delta t$ of the standard fields during the night and the colour coverage $\Delta C$ of the observed standard stars. The data were considered adequate if $\Delta C \geq 1$~mag and $\Delta t \geq 4$~hrs. Then, we used the uncertainties on $ZP_M$ to decide upon their quality: if $\delta ZP_M \leq 0.25$~mag in all bands, the data were considered good, if $\delta ZP_M \leq 0.25$~mag for one band only or less, the data were considered bad, and in all remaining cases the data were considered of intermediate quality.

\begin{table}
\centering
\caption{Calibrated magnitudes for the  198 good SPSS candidates. The full table is available online and at the CDS.
\label{tab:phot}}
\begin{tabular}{lll}
\hline
Column & Units & Description \\
\hline
SPSS ID & --- & The internal SPSS ID (001-399) \\
SPSS name & --- & The SPSS adopted name \\
R.A. (J2000) & hh:mm:ss.ss & Right ascension$^{\ddagger}$ \\
Dec (J2000) & dd:mm:ss.s & Declination$^{\ddagger}$  \\
$B$ & mag & $B$ calibrated magnitude \\
$V$ & mag & $V$ calibrated magnitude \\
$R$ & mag & $R$ calibrated magnitude \\
$I$ & mag & $I$ calibrated magnitude \\
$\delta B$ & mag & uncertainty on $B$ \\
$\delta V$ & mag & uncertainty on $V$ \\
$\delta R$ & mag & uncertainty on $R$ \\
$\delta I$ & mag & uncertainty on $I$ \\
N$_{\rm{B}}$ & --- & number of nights for $B$ \\
N$_{\rm{V}}$ & --- & number of nights for $V$ \\
N$_{\rm{R}}$ & --- & number of nights for $R$ \\
N$_{\rm{I}}$ & --- & number of nights for $I$ \\
P$_{\rm{B}}$ & --- & number of measurements for $B$ \\
P$_{\rm{V}}$ & --- & number of measurements for $V$ \\
P$_{\rm{R}}$ & --- & number of measurements for $R$ \\
P$_{\rm{I}}$ & --- & number of measurements for $I$ \\
Flag$_B$ & --- & 1 if no fully photometric nights \\
Flag$_V$ & --- & 1 if no fully photometric nights \\
Flag$_R$ & --- & 1 if no fully photometric nights \\
Flag$_I$ & --- & 1 if no fully photometric nights \\
Notes & --- & Annotations$^*$\\
\hline                                                     \end{tabular}
\\ 
$^{\ddagger}$ - \cite{PANCINO2012} and references therein.\\ 
$^*$Including the availability of photometry in the Landolt or Clem collections and of CALSPEC flux tables.
\end{table}

At that point, we judged a night as {\em photometric} if the Landolt data were adequate to compute $k_M$ and $ZP_M$ and if the quality of their determination was good. The night was judged not useful if the data are not adequate to calibrate it (unknown quality) or if they were adequate but provided bad $k_M$ or $ZP_M$ estimates for at least two bands (not photometric). In all other cases the night was judged as relatively good, i.e. still useful to compute  absolutely calibrated magnitudes, but to be used with extra care.

\subsection{Final {\em absolutely calibrated} magnitudes}

Once the coefficients $k_M$, $f_M$, and $ZP_M$ of Equations~\ref{eq:am},\ref{eq:amkb} and \ref{eq:sol}  were determined using Landolt fields observations, the same equations could be used to obtain $M$ from $m$ and the known coefficients, at least for the nights that passed the QC. For each frame catalogue of aperture magnitudes that passed the QC procedures, we thus obtained absolutely calibrated magnitudes, with their own propagated uncertainties and quality flags (in case the QC was only partially passed). 

\begin{figure}
\includegraphics[clip,width=\columnwidth]{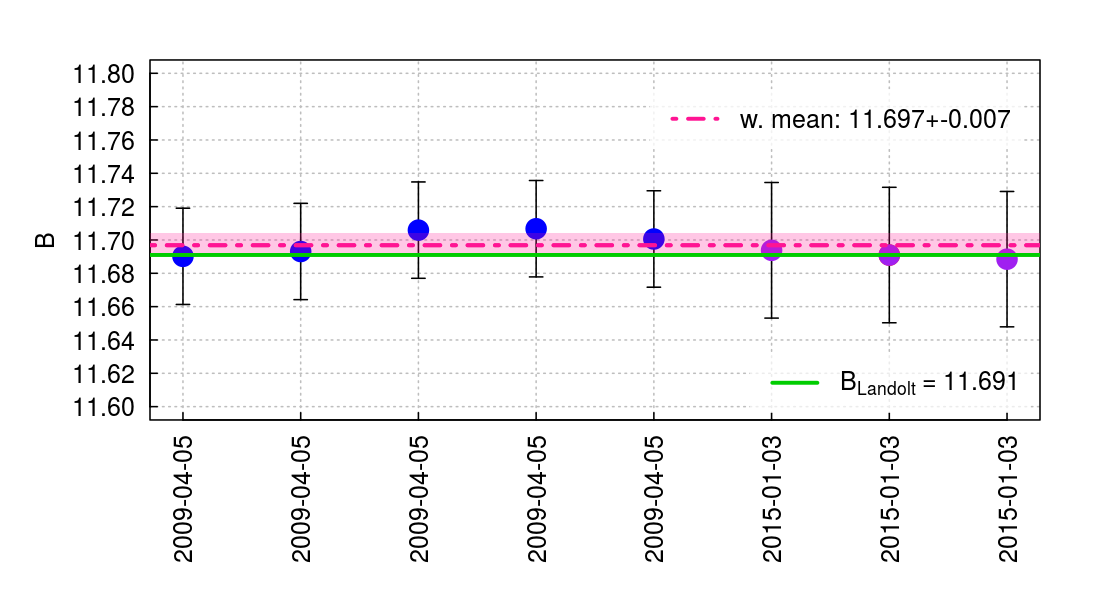}
\includegraphics[clip,width=\columnwidth]{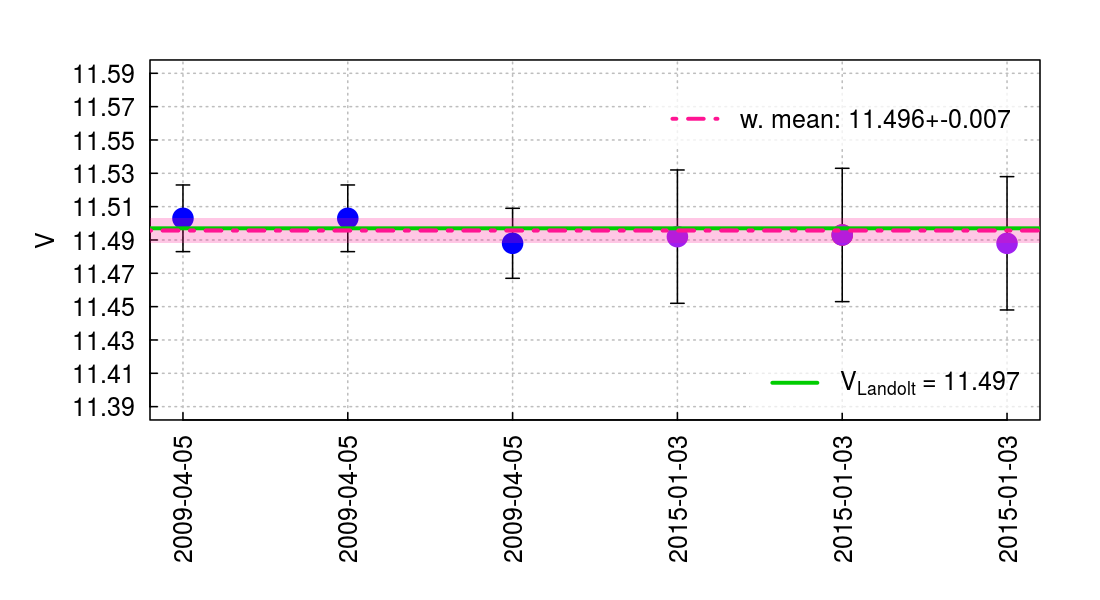}
\includegraphics[clip,width=\columnwidth]{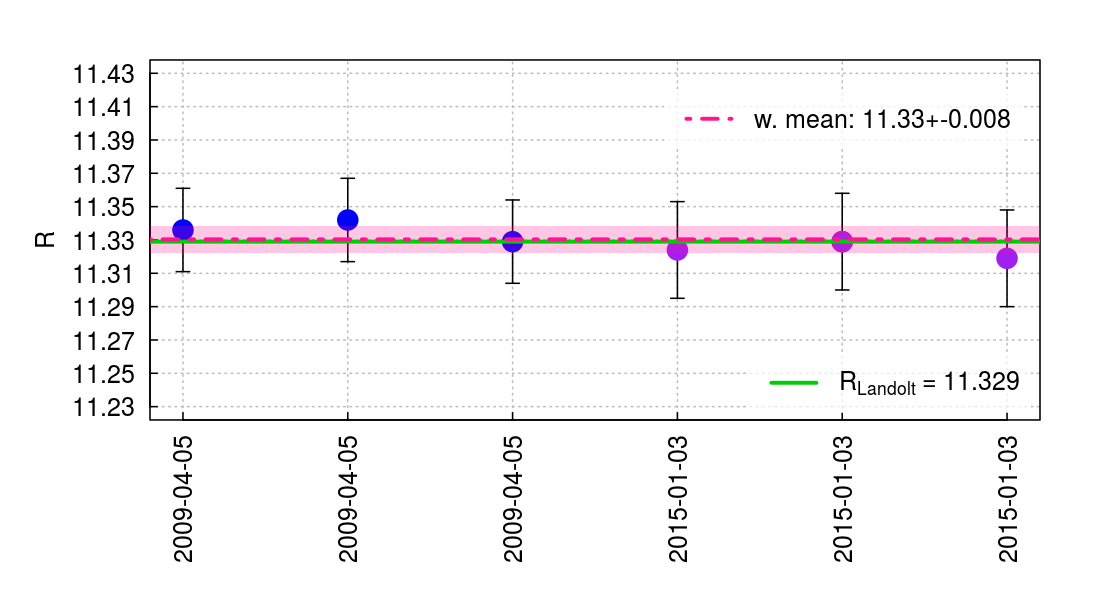}
\caption{Example of checkplots for the computation of the combined calibrated magnitude of SPSS\,012 (LTT\,4364). 
Each panel shows the magnitude measured in a given band, as indicated, versus the
correponding date.
Each observation is plotted as a blue circle in case of photometric nights (Section~\ref{sec:qc}) or a purple one in case of intermediate quality nights. The weighted average is reported as a red line with a shaded strip representing the error, while the Landolt magnitude (Section~\ref{sec:valid}) as a green line.} \label{fig:med}
\end{figure}

We then computed the weighted mean and the corresponding weighted standard deviation of the epoch calibrated magnitudes to obtain the final calibrated $BVR(I)$ magnitudes for each SPSS. An example of the type of datasets in hand is presented in Figure~\ref{fig:med}. In order to reject a handful of very obvious {\em extreme outliers} from the weighted mean computation, we first used all data available in a given band for a given SPSS to compute an initial weighted mean, then we rejected all points deviating more than 0.5 mag if $\sigma < 0.5$ mag, or all points deviating more than $1\sigma$ if $\sigma > 0.5$ mag, and finally we re-computed the final weighted mean with the surviving points. The above thresholds were chosen after experimenting with the data, with the goal of removing only measurements that are clearly very different from the remaining ones.

\begin{figure}
\includegraphics[width=\columnwidth]{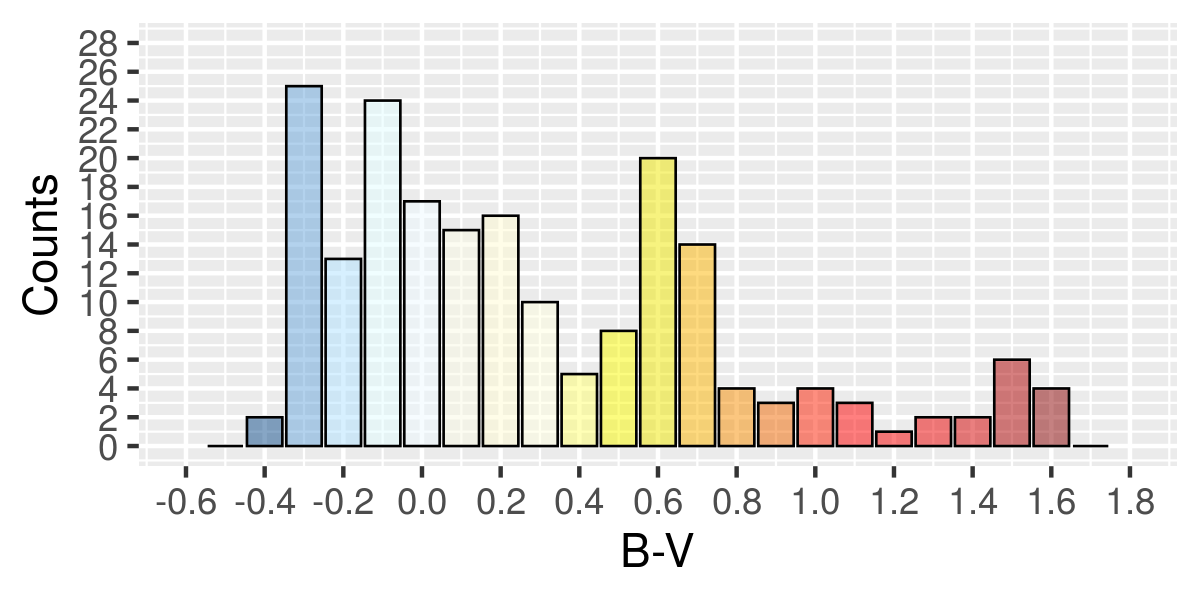} 
\includegraphics[width=\columnwidth]{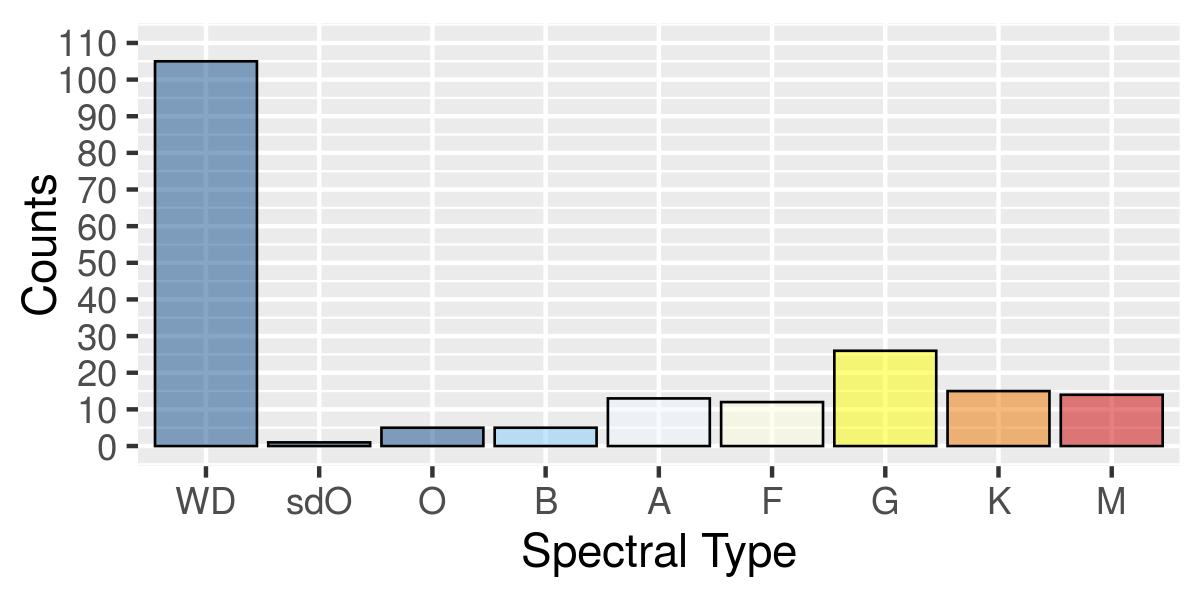} 
\includegraphics[width=\columnwidth]{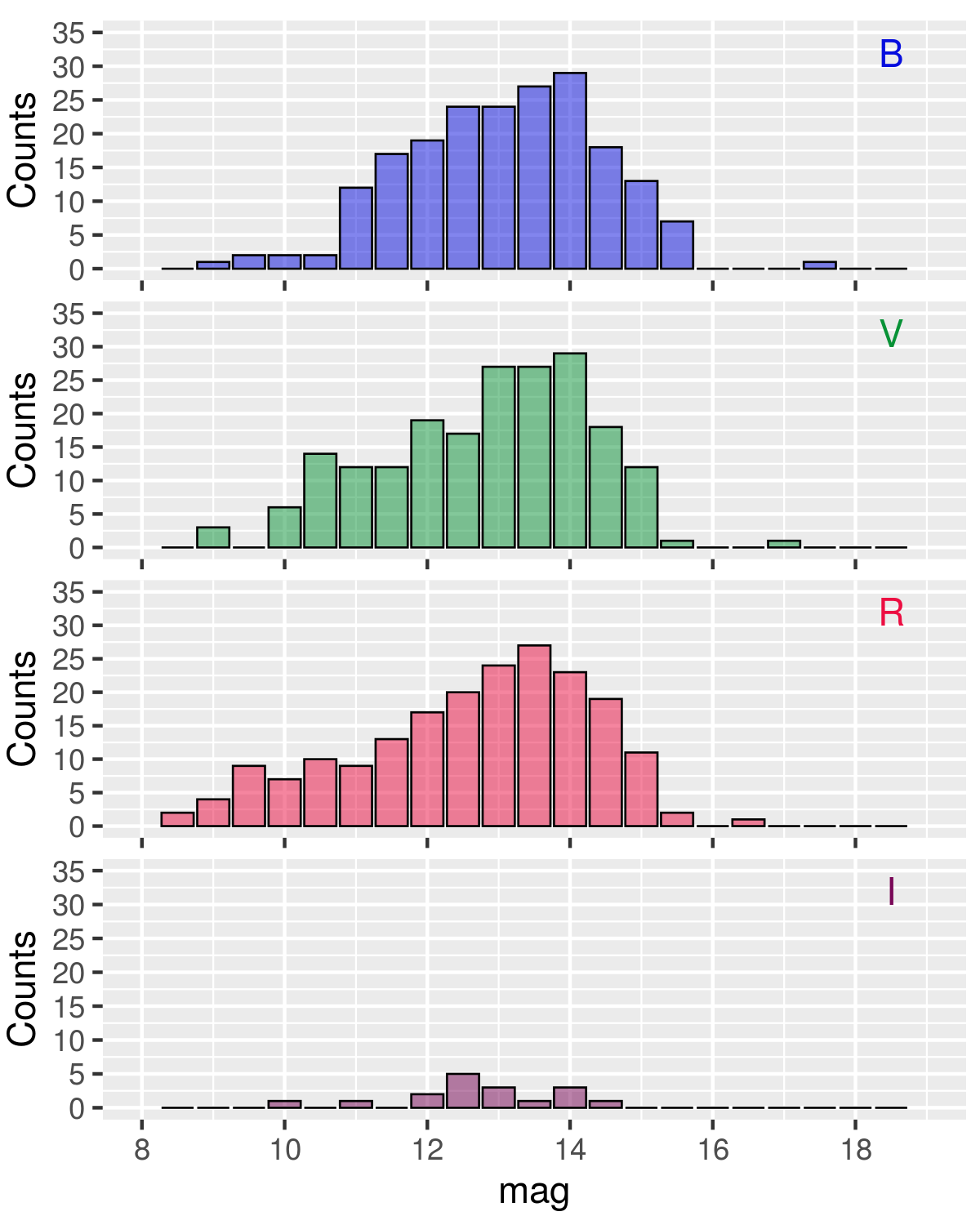}
\caption{Distribution of the SPSS candidates as a function of colour  and spectral type (multi-coloured histograms, top panels) and of the $BVRI$ magnitudes (single-colour histograms, remaining panels).}
\label{fig:hist}   
\end{figure}

We also noted that when data for a certain SPSS came from a single night, the uncertainty resulting from the weighted average was  unrealistically  small, because all measurements were very similar to each other. This was not the case when really independent measurements were available, from different observing nights and different facilities. For this reason, when  only one observing night was available, we also computed the median of the errors of the individual measurements.  The final error for these cases was chosen as the largest between this and the error on the weighted mean. 

\begin{table*}
\centering
\tabcolsep 5pt
\caption{Calibrated magnitudes ($M$) and uncertainties ($\delta M$) of rejected SPSS candidates, with motivation of the rejection. }\label{tab:rej}
\begin{tabular}{|l|l|r|r|r|r|r|r|r|r|r|r|r|l|}
\hline 
SPSS ID & SPSS name & RA (J2000) & Dec (J2000)  & $B$ & $V$  & $R$ & $I$ & $\delta B$ & $\delta V$ & $\delta R$ & $\delta I$ & Notes \\
& & (hh:mm:ss) & (dd:mm:s) & (mag) & (mag) & (mag) & (mag) & (mag) & (mag) & (mag) & (mag) & \\
\hline  
018 & BD +284211  & 21:51:11.02$^{\ddagger}$ & +28:51:50.4$^{\ddagger}$ &  10.202 & 10.539 & 10.695 & 10.893 &  0.022  & 0.025 & 0.013 & 0.023 & *,$\dagger$,1 \\ 
020 & BD +174708  & 22:11:31.37$^{\ddagger}$ & +18:05:34.2$^{\ddagger}$ &   9.895 &  9.474 &  9.164 &  8.853 &  0.012  & 0.009 & 0.031 & 0.024 & *,$\dagger$,2 \\      
029 & SA 105-448  & 13:37:47.07$^{\ddagger}$ & -00:37:33.0$^{\ddagger}$  &   9.386 &  9.149 &  8.998 &      - &  0.033  & 0.035 & 0.022 &     - & *,3 \\      
034 & 1740346     & 17:40:34.68$^{\ddagger}$ & +65:27:14.8$^{\ddagger}$ &  12.745 & 12.578 & 12.473 &      - &  0.024  & 0.016 & 0.012 &     - & 3,$\dagger$ \\        
051 & HD 37725    & 05:41:54.37$^{\ddagger}$ & +29:17:50.9$^{\ddagger}$ &   8.501 &  8.328 &  8.340 &      - &  0.021  & 0.011 & 0.125 &     - & 3,$\dagger$ \\ 
131 & WD 1327-083 & 13:30:13.64$^{\ddagger}$ & -08:34:29.5$^{\ddagger}$ &  12.395 & 12.335 & 12.414 &      - &  0.017  & 0.019 & 0.021 &     - & 1 \\ 
150 & WD 2126+734 & 21:26:57.70$^{\diamond}$ & +73:38:44.4$^{\diamond}$ &  12.888 & 12.863 & 12.910 &      - &  0.026  & 0.017 & 0.024 &     - & 4 \\  
152 & G 190-15    & 23:13:38.82$^{\ddagger}$ & +39:25:02.6$^{\ddagger}$ &  11.685 & 11.051 & 10.606 & 10.192 &  0.018  & 0.046 & 0.038 & 0.046 & 5 \\  
160 & WD0104-331  & 01:06:46.86$^{\ddagger}$ & -32:53:12.4$^{\ddagger}$ &  18.232 & 17.051 & 16.292 &      - &  0.030  & 0.031 & 0.034 &     - & 6 \\
190 & G 66-59     & 15:03:49.01$^{\diamond}$ & +10:44:23.3$^{\diamond}$ &  13.867 & 13.167 & 12.763 &      - &  0.042  & 0.046 & 0.036 &     - & 7 \\  
212 & WD 2256+313 & 22:58:39.44$^{\ddagger}$ & +31:34:48.9$^{\ddagger}$ &  -      & 13.988 &  -     & 11.803 &      -  & 0.054 &     - & 0.071 & 8 \\ 
216 & WD 0009+501 & 00:12:14.80$^{\ddagger}$ & +50:25:21.4$^{\ddagger}$ &  14.812 & 14.373 & 14.090 & 13.785 &  0.004  & 0.011 & 0.014 & 0.044 & 1 \\     
227 & WD 0406+592 & 04:10:51.70$^{\ddagger}$ & +59:25:05.0$^{\ddagger}$ &  14.511 & 14.716 & 14.809 &      - &  0.025  & 0.009 & 0.025 &     - & 9 \\     
232 & LP 845-9    & 09:00:57.23$^{\diamond}$ & -22:13:50.5$^{\diamond}$ &  16.196 & 14.623 &      - &      - &  0.038  & 0.046 &     - &     - & 8 \\     
240 & WD 1121+508 & 11:24:31.41$^{\diamond}$ & +50:33:31.6$^{\diamond}$ &  15.601 & 14.910 & 14.516 &      - &  0.031  & 0.040 & 0.019 &     - & 8 \\          
243 & WD 1232+479 & 12:34:56.20$^{\diamond}$ & +47:37:33.4$^{\diamond}$ &  14.832 & 15.325 & 14.556 &      - &  0.472  & 0.293 & 0.028 &     - & 10 \\           
265 & WD 2010+310 & 20:12:22.28$^{\diamond}$ & +31:13:48.6$^{\diamond}$ &  -      &  14.833 & 14.821 & 15.015 &     -  & 0.031 & 0.022 & 0.040 & 8 \\ 
286 & WD 0205-304 & 02:07:40.71$^{\diamond}$ & -30:10:57.6$^{\diamond}$ &  15.732 & 15.750 & 15.826 &      - &  0.024  & 0.019 & 0.017 &     - & 8 \\   
290 & WD 1230+417 & 12:32:26.18$^{\diamond}$ & +41:29:19.2$^{\diamond}$ &  15.644 & 15.726 & 15.818 &      - &  0.024  & 0.028 & 0.014 &     - & 8 \\   
293 & WD 1616-591 & 16:20:34.75$^{\diamond}$ & -59:16:14.2$^{\diamond}$ &  15.105 & 14.998 & 15.034 &      - &  0.023  & 0.006 & 0.004 &     - & 11 \\ 
294 & WD 1636+160 & 16:38:40.40$^{\diamond}$ & +15:54:17.0$^{\diamond}$ &  15.744 & 15.652 & 15.700 &      - &  0.028  & 0.032 & 0.015 &     - & 8 \\   
297 & PG 0924+565 & 09:28:30.52$^{\diamond}$ & +56:18:11.6$^{\diamond}$ &  16.540 & 16.464 & 16.079 &      - &  0.189  & 0.252 & 0.017 &     - & 8 \\   
316 & SDSS13028   & 16:40:24.18$^{\ddagger}$ & +24:02:14.9$^{\ddagger}$ &  15.510 & 15.275 & 15.120 &      - &  0.015  & 0.012 & 0.019 &     - & 3 \\
317 & SDSS15724   & 20:47:38.19$^{\ddagger}$ & -06:32:13.1$^{\ddagger}$ &  15.354 & 15.125 & 14.946 &      - &  0.292  & 0.207 & 0.168 &     - & 12 \\  
318 & SDSS14276   & 22:42:04.17$^{\ddagger}$ & +13:20:28.6$^{\ddagger}$ &  14.515 & 14.313 & 14.204 &      - &  0.051  & 0.052 & 0.041 &     - & 3 \\
322 & SDSS12720   & 12:22:41.66$^{\ddagger}$ & +42:24:43.7$^{\ddagger}$ &  15.526 & 15.372 & 15.136 &      - &  0.252  & 0.175 & 0.154 &     - & 3 \\     
325 & SDSS03932   & 00:07:52.22$^{\ddagger}$ & +14:30:24.7$^{\ddagger}$ &  15.401 & 15.086 & 14.880 &      - &  0.018  & 0.021 & 0.020 &     - & 3 \\    
329 & GJ 207.1    & 05:33:44.81$^{\diamond}$ & +01:56:43.4$^{\diamond}$ &  13.066 & 11.523 & 10.447 &      - &  0.020  & 0.024 & 0.024 &     - & 13\\ 
330 & GJ 268.3    & 07:16:19.77$^{\diamond}$ & +27:08:33.1$^{\diamond}$ &  12.403 & 10.875 &  9.810 &      - &  0.005  & 0.004 & 0.009 &     - & 14 \\    
350 & LTT377      & 00:41:30.47$^{\ddagger}$ & -33:37:32.0$^{\ddagger}$ &  12.000 & 10.565 &  9.640 &      - &  0.010  & 0.009 & 0.027 &     - & 15 \\  
\hline
\end{tabular}
\\
$^{\ddagger}$ - \cite{PANCINO2012} and references therein;
$^{\diamond}$ - 2MASS  All-Sky Catalog of Point Sources \citep{CUTRI2003};
* - Literature data from Landolt available (see Section~\ref{sec:valid});
$\dagger$ - CALSPEC spectrum available  (see Section~\ref{sec:valid});
1 - suspected variable \citep{MARINONI2016};
2 - long term variability  \citep{MARINONI2016, BOHLINLANDOLT2015};
3 - variable \citep{MARINONI2016};
4 - binary star \citep{FARIHI2005};
5 - suspected long term variable \citep{MARINONI2016};
6 - discordant coordinates in \citealt{WEGNER1973} and \citealt{CHAVIRA1958};
7 - double-lined spectroscopic binary  \citep{LATHAM1992};
8 - too faint;
9 - close red companions;
10 - radial velocity variable  \citep{SAFFER1998};
11 - high PM star, now too close to another faint star;
12 - variable star \citep{ABBAS2014};
13 - Wachmann variable, flare Star \citep{LUYTEN1954}; 
14 - spectroscopic binary  \citep{JENKINS2009};
15 - suspected variable \citep{SAMUS2017, JAYASINGHE2019}.
\end{table*}

\section{Results and validation}
\label{sec:valid}

The final calibrated magnitudes for  the 198 good SPSS candidates are listed in Table~\ref{tab:phot}, which is available electronically. The magnitude measurements for  30 SPSS that were rejected or do not have spectra of sufficient quality (see Section~\ref{sec:obs}) are presented separately, in Table~\ref{tab:rej}, along with the motivation for the rejection. The distributions of the entire sample in magnitude and colour are presented in Figure~\ref{fig:hist}, where the large colour baseline and magnitude range can be fully appreciated. In the following, we compare our measurements with various literature sources, most notably the collection of photoelectric and CCD photometry by Landolt (Section~\ref{sec:landolt}) and the spectra of the CALSPEC collection (Section~\ref{sec:calspec}), finding a good agreement within the uncertainties, i.e., of the order of 1 per cent.

\begin{figure*}
\includegraphics[width=\columnwidth]{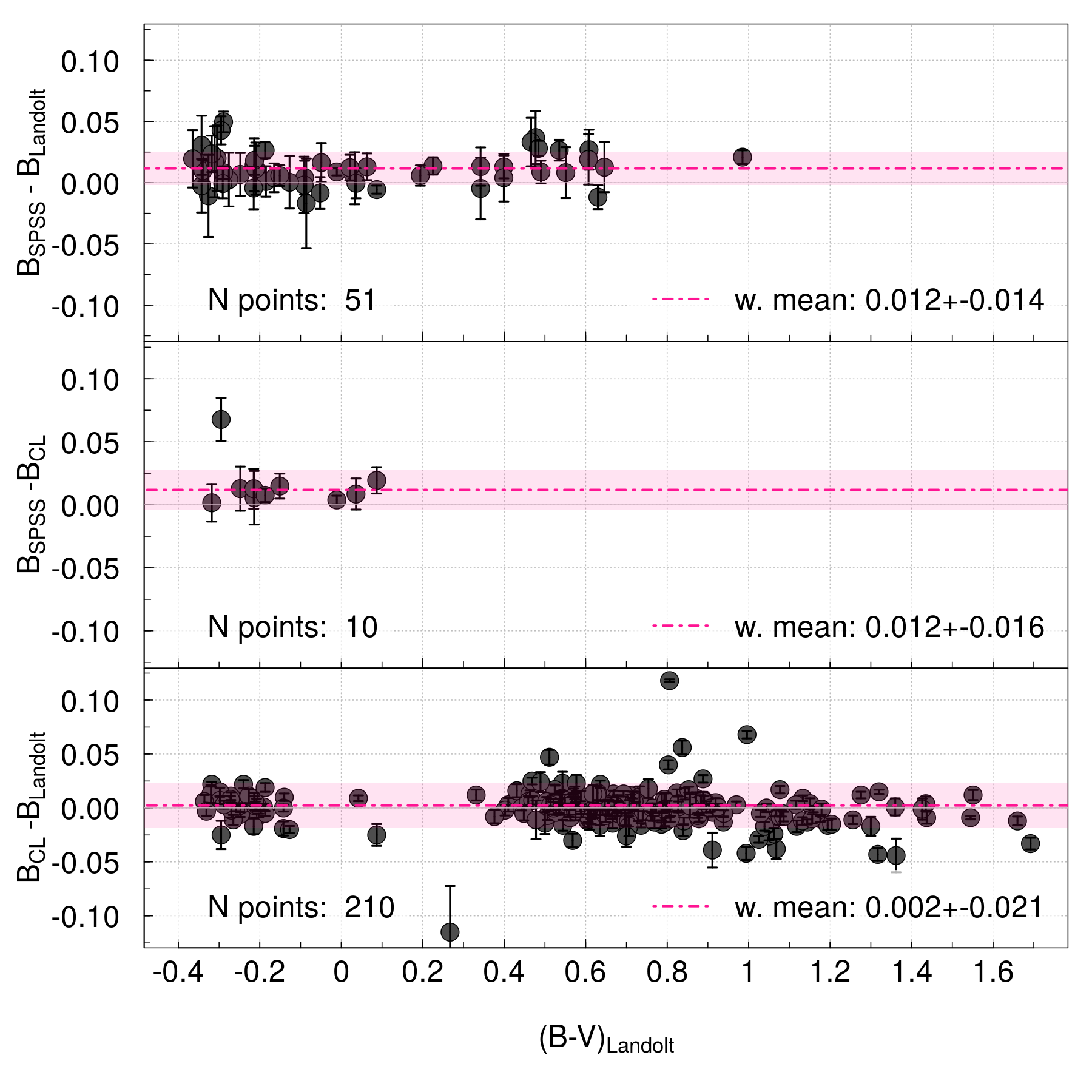}
\includegraphics[width=\columnwidth]{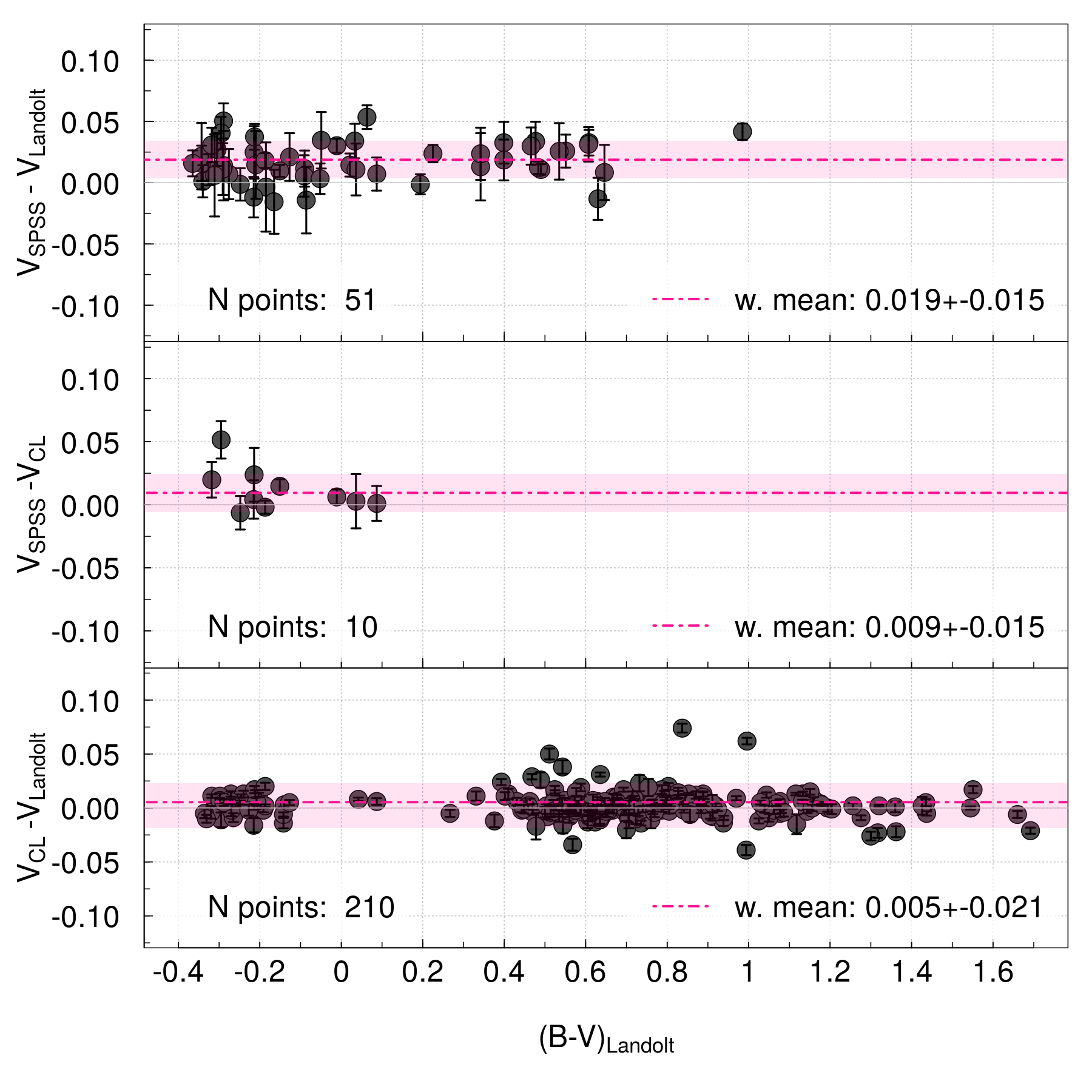}
\includegraphics[width=\columnwidth]{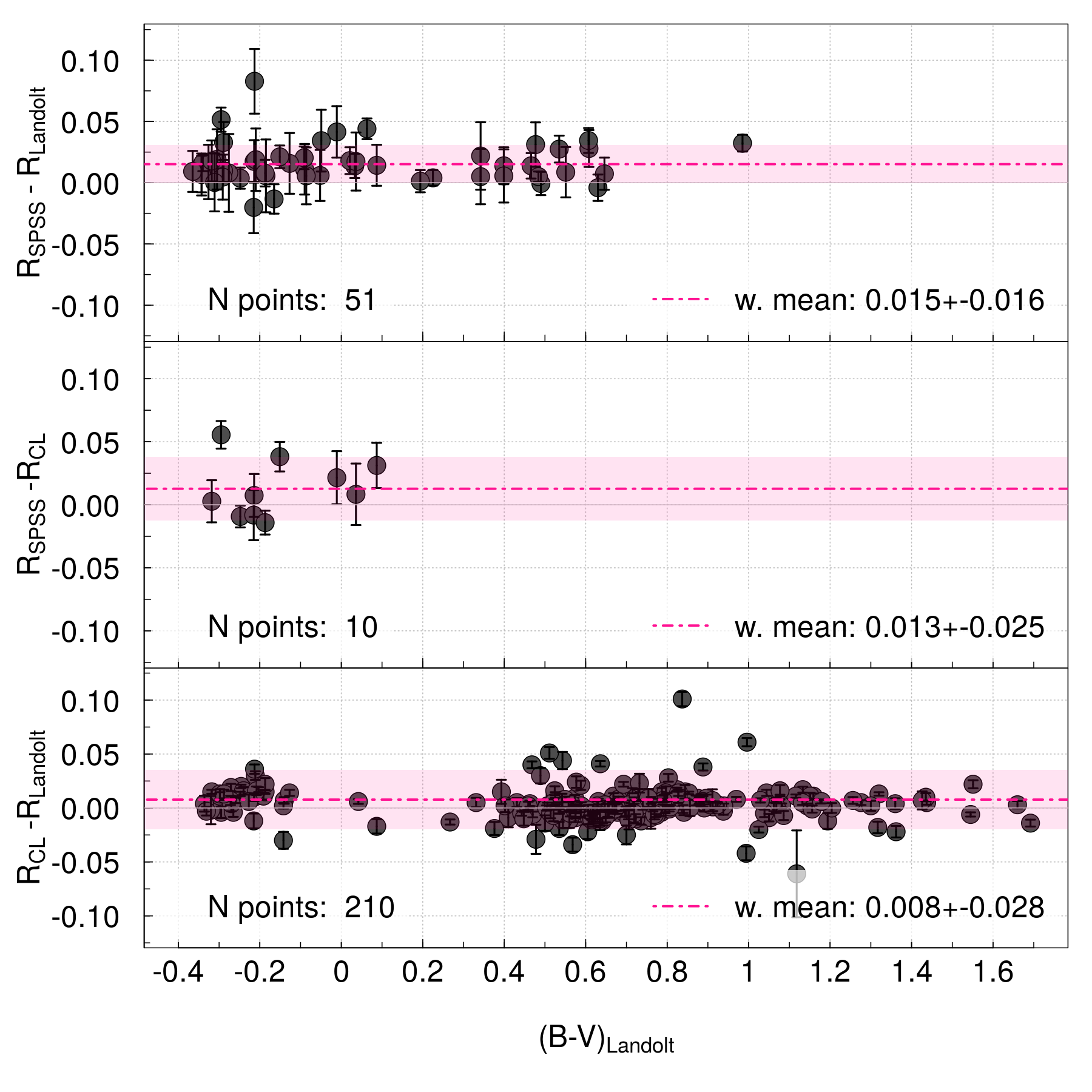}
\includegraphics[width=\columnwidth]{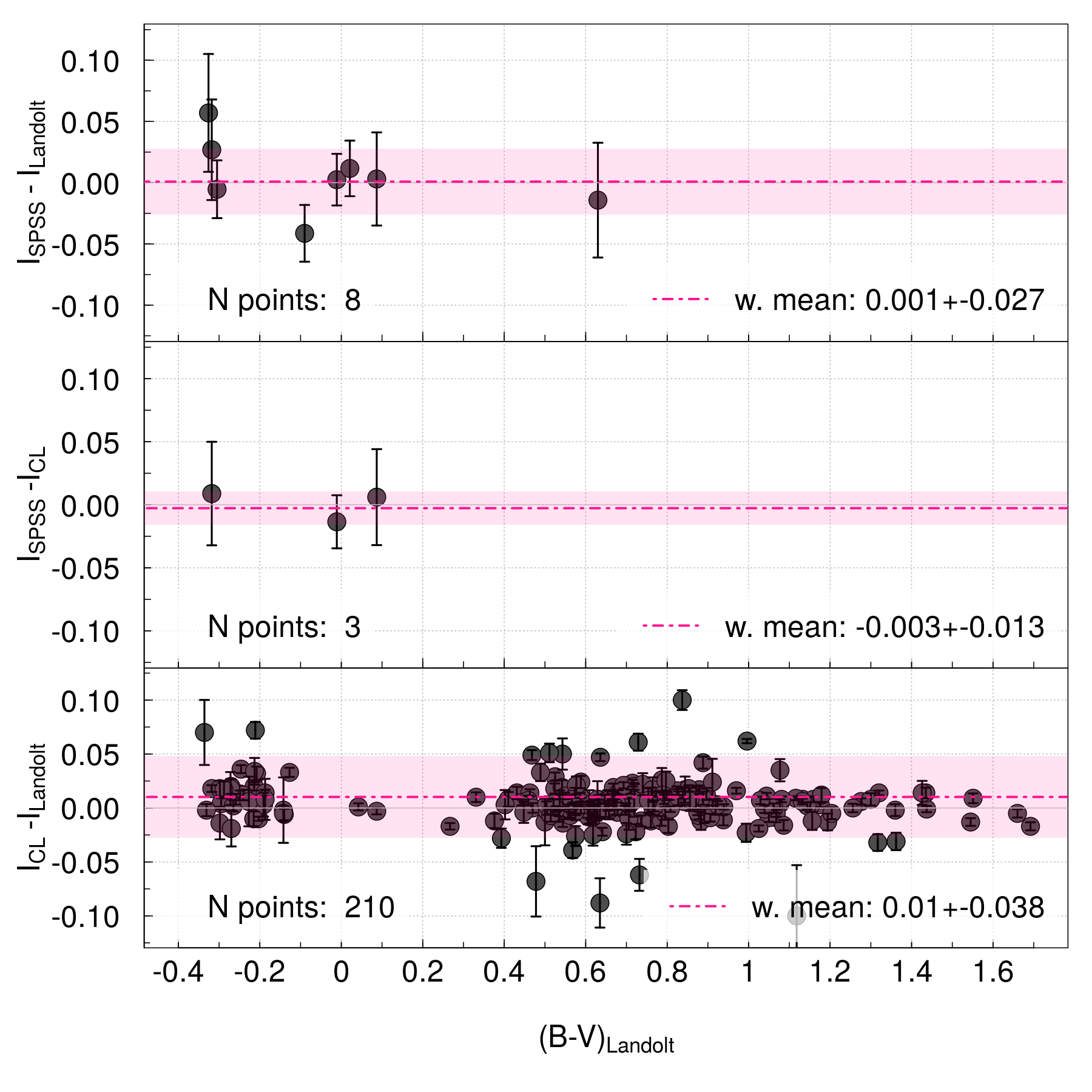}
\caption{Comparison among the magnitudes of the SPSS, the Landolt, and the Clem  collection (see Section~\ref{sec:landolt} for details). Each of the four panels shows the results of one photometric band as a function of the $B-V$ colour. In each panel, the top graph shows the comparison between the SPSS and Landolt collections, the middle one between the SPSS and Clem collections, and the bottom one between the Clem and Landolt collections. The weighted averages are reported in each sub-panel, along with the number of stars in common between each pair of collections.  One star lies outside the vertical limits in the bottom graph of the $BVR$ band panels and two stars lie outside the limits in  the bottom graph of the $I$ band panel because we used the same limits as in Fig.~\ref{fig:DU13menoCalspec}}.
\label{fig:landolt}   
\end{figure*}

\subsection{Comparison with Landolt photometry}
\label{sec:landolt}

While our data are calibrated on the original set of photometric standard stars defined by \citet{LANDOLT1992a}, there are several papers by the same author and collaborators, providing photoelectric and CCD magnitudes reported as accurately as possible on the original photoelectric system by \citet{LANDOLT1992a}. The collection is useful to validate our work, thus we selected measurements from \citet{LANDOLT1973,LANDOLT1983,LANDOLT1992a,LANDOLT1992b}, \citet{LANDOLTUOMOTO2007}, \citet{LANDOLT2007,LANDOLT2009,LANDOLT2013}, and \citet{BOHLINLANDOLT2015}. Small inconsistencies and non-linear relations between different instruments and especially between photoelectric and CCD magnitudes were corrected by the  listed authors with a series of {\em ad hoc} transformations and piecewise polynomials, depending on the dataset \citep[see, e.g.,][]{LANDOLT2007}. The internal consistency of the entire set of magnitudes was estimated to be of the order of 1 per cent. We merged the data by keeping the most recent measurement available for each star, and created a unique catalogue  with $\simeq$1650 stars, that we will refer to in the following as ``Landolt collection".  

We excluded from any comparison the rejected candidate SPSS (see Table~\ref{tab:rej}); the final sample of SPSS in common with the Landolt collection contains 51 stars (the stars used are labelled in Table~\ref{tab:phot}). The results of the comparison are shown in Figure~\ref{fig:landolt}, in the top sub-panels of each of the $B$, $V$, $R$, and $I$ panels. As it can be seen, while the mean offsets are generally smaller than the uncertainties, there are offsets of about 1 per cent between our SPSS magnitudes and the ones in the Landolt collection. The offset is consistently positive, i.e., our magnitudes are consistently fainter than the ones in the Landolt collection, but it varies a lot from one band to the other. For example, the $I$ band has a 0.001 offset while the $V$ band has an offset of 0.019~mag. Although contained within uncertainties, these offsets are non-negligible compared to our original requirement on the SPSS flux tables (spectra calibrated to 1--3 per cent) for the external flux calibration of {\em Gaia} data.

We thus considered another useful photometric dataset, published by \citet{ClemLandolt2013,ClemLandolt2016}, to look further into this matter. The full dataset, that we will refer to as ``Clem collection", contains a much larger number of sources than the Landolt collection, but only 10 sources in common with our SPSS set. Their magnitudes were calibrated as accurately as possible onto the \citet{LANDOLT1992a} set and on the full Landolt collection, using {\em ad hoc} techniques similar to the ones used in the Landolt collection. As shown by \citet{LANDOLT2007} before, the basic problem is that the diversity of detectors and filters employed in the various studies implies a certain minimum degree of uncertainty \citep[see][for an in-depth discussion]{ClemLandolt2013}, that can be quantified as being of the order of 1 per cent with current technology. The corrections applied by \citet{ClemLandolt2013,ClemLandolt2016} bring their measurements onto the original Landolt system to about 0.5 per cent on average across the five bands. We show this comparison in the bottom sub-panels of Figure~\ref{fig:landolt} for each of four bands used in this study. As it can be seen, the Clem collection is also consistently fainter than the Landolt collection, similarly to what happens to the SPSS magnitudes, but with a smaller offset: on average $\simeq$0.5 per cent rather than $\simeq$1 per cent, depending on the band. As a result, the SPSS magnitudes agree generally slightly better with the Clem collection than with the Landolt one, as shown in the middle sub-panels of Figure~\ref{fig:landolt}. We observe from Figure~\ref{fig:landolt} that the spread of our measurements is generally comparable to or even smaller than the spreads of the Landolt and Clem samples, perhaps because of the careful exclusion of variable stars with amplitude larger than $\pm$5~mmag from our sample \citep{MARINONI2016}.

\subsection{Comparison with CALSPEC spectrophotometry} 
\label{sec:calspec}

Another fundamental reference set for flux calibrations is the CALSPEC\footnote{\url{https://www.stsci.edu/hst/instrumentation/reference-data-for-calibration-and-tools/astronomical-catalogs/calspec}} set of spectrophotometric standards \citep{bohlin14}. It  currently contains about 95 stars with complete coverage of the {\em Gaia} wavelength range (and beyond), observed with the Hubble Space Telescope (HST), and calibrated in flux to an  accuracy of about 1 per cent (on the {\it reference system}), although recent revisions altered the flux of {\it individual stars} by up to 2.5 per cent  \citep{bohlin19}. All spectra are calibrated on three primary calibrators, the pure hydrogen white dwarfs GD\,71, GD\,153, and G\,191--B2B and on theoretical spectra \citep{bohlin17}.

\begin{figure}
\includegraphics[width=\columnwidth]{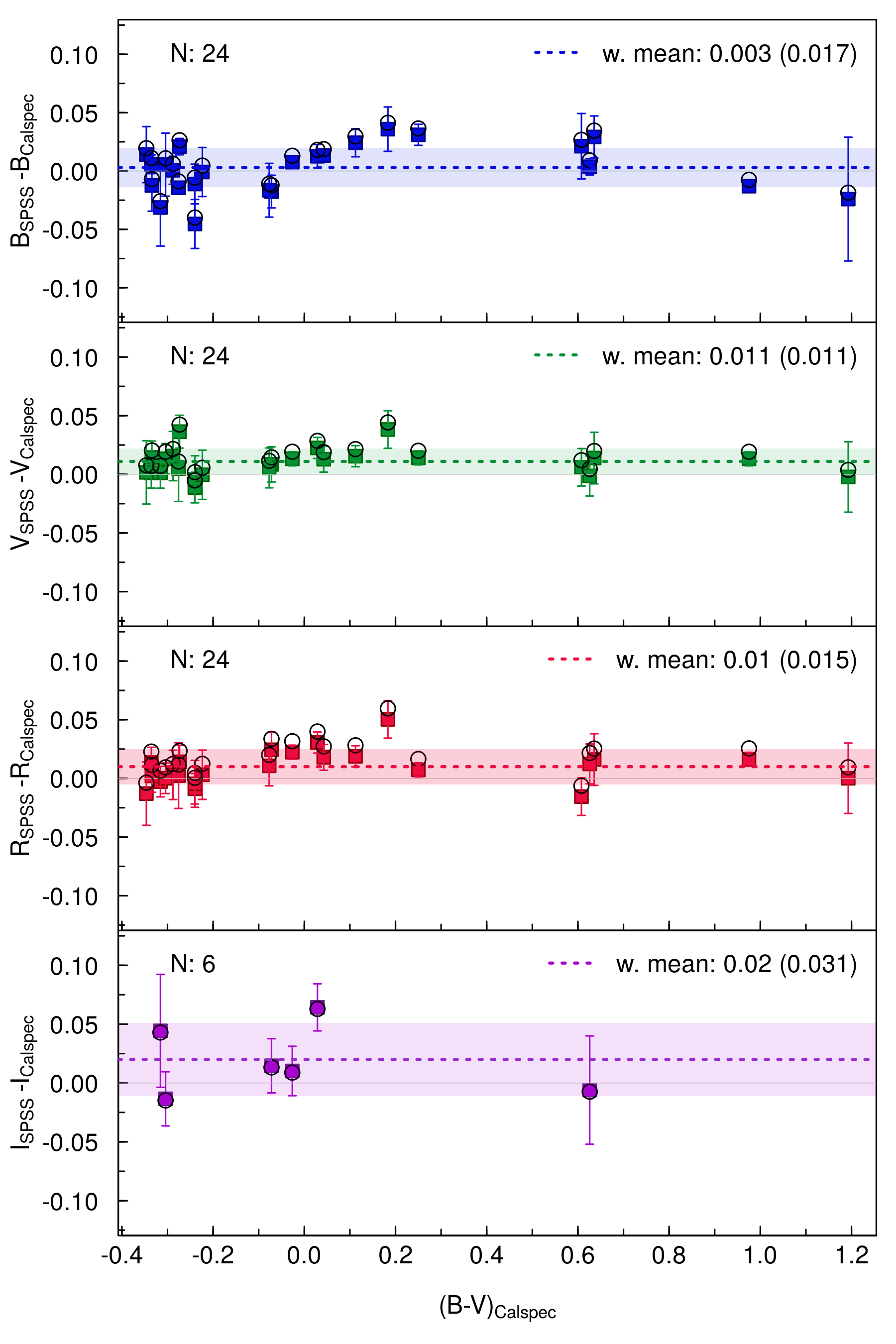}
\caption{Comparison between the SPSS photometry and the synthetic photometry computed on the corresponding CALSPEC spectra using the passbands by \citet{BESSELL2012}.
Filled and empty symbols correspond to the synthetic photometry computed using as reference star the Vega spectrum by CALSPEC or the one adopted for the Gaia photometric system respectively  (see Sect.~\ref{sec:calspec}).
Each panel represents a different passband, as annotated, and the weighted mean of the differences  (computed on the filled symbols)  is reported in each panel and marked with a dotted line. }
\label{fig:DU13menoCalspec}
\end{figure}
\begin{figure}
\includegraphics[width=\columnwidth]{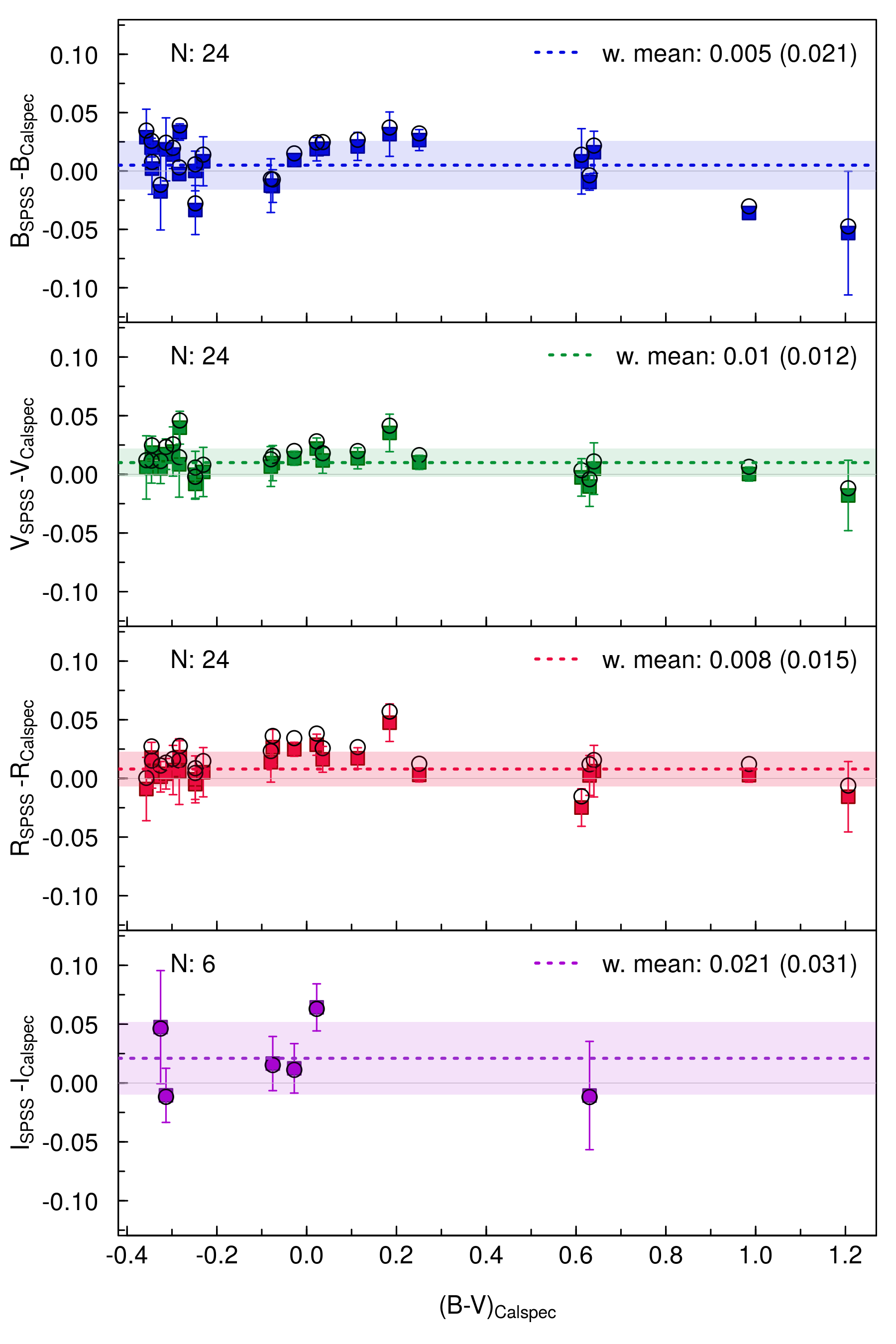} 
\caption{Same as Figure~\ref{fig:DU13menoCalspec}, but applying the wavelength shifts proposed by \citet{BOHLINLANDOLT2015} to the \citet{BESSELL2012} passbands.}
\label{fig:DU13menoCalspecwithshift}
\end{figure}

We computed synthetic magnitudes for 24 CALSPEC spectra that correspond to good SPSS candidates. We only considered CALSPEC spectra covering the entire {\em Gaia} wavelength range, and we did not include rejected SPSS (Table~\ref{tab:rej}) in the comparison (the stars used are labelled in Table~\ref{tab:phot}). We selected the latest and most complete spectra in the CALSPEC database at the time of writing   plus the model spectra for the three CALSPEC primary  calibrators \citep{BOHLIN2020b}. We used the passbands by \citet{BESSELL2012} and as reference spectrum the same high-fidelity Kurucz model of Vega used by CALSPEC (alpha\_lyr\_mod\_004)  and Vega magnitudes given by \cite{CASAGRANDE2014} and references therein. We then compared the synthetic CALSPEC magnitudes with our measurements, as illustrated in Figure~\ref{fig:DU13menoCalspec}. Our measurements agree with the synthetic photometry obtained from the CALSPEC spectra within 1-2 per cent. In the $B$ band, a colour dependence is apparent, in the form of a curved trend with a maximum difference around $B-V=$ 0.0--0.3~mag. A similar but less pronounced trend can be observed for the $V$ and $R$ bands, while the $I$ band data are too sparse to draw any conclusion. These trends were not present in the comparisons with the Landolt and Clem collections, but they were observed by \citet{BOHLINLANDOLT2015} in their comparison of CALSPEC synthetic photometry with Landolt photometry. They mitigated the trend by applying small wavelength shifts to the \citet{BESSELL2012} passbands used to obtain their CALSPEC photometry, which are the same that we used. A similar procedure was applied by \citet{STRITZINGER2005} to remove similar colour trends in the comparison between their synthetic magnitudes and literature data. Along those lines, \citet{BESSEL1990} suggested that "optimum passbands" may be recovered by minimizing the differences between synthetic and standard magnitudes. It is extremely interesting that \citet{ClemLandolt2013,ClemLandolt2016} also found similar colour trends in their comparison with the original Landolt collection, that they corrected with empirical methods, as described in the previous section.

To test the idea, we repeated the comparison, this time applying the same wavelength shifts employed by \citet{BOHLINLANDOLT2015} to the \citet{BESSELL2012} passbands. The applied shifts were $-20$~\AA\ for the $B$ and the $V$ bands, $-31$~\AA\ for the $R$ band, and $-27$~\AA\ for the $I$ band. The results are shown in Figure~\ref{fig:DU13menoCalspecwithshift}. As it can be observed, the residuals becomes flatter in the colour range $-0.4  \lesssim B-V  \lesssim 0.6$, i.e. the one covered by \citealt{BOHLINLANDOLT2015}), but the comparison gets significantly worse for redder objects. The worsening of the comparison for red sources is confirmed by the $V$ and $R$ bands results. For this reason, we think that a wavelength shift of the passbands is not sufficient to fully repair the discrepancy. The results in Figures~\ref{fig:DU13menoCalspec},\ref{fig:DU13menoCalspecwithshift} do not change significantly
using as reference spectrum for the synthetic photometry the one chosen for the Gaia photometric system \citep{CARRASCO2016}, i.e. a Kurucz/ATLAS9 model \citep{BUSER1992} for the Vega spectrum (CDROM 19) with $T_{eff}=9400$ K, log~$g = 3.95$ dex, [M/H]=-0.5 dex, and vmicro = 2 km s$^{-1}$,  scaled to fit STIS data (over the interval 5545-5570 \AA) \citep{BOHLIN2004}.

As extensively discussed in the cited literature and in the previous section, the original Landolt system was defined using a 1P21 photomultiplier \citep{LANDOLT1973}, while the following literature used a variety of systems, including different photomultipliers, CCD cameras, optical filters, telescopes, and detectors. The original Landolt system provided a standard  system that is used to calibrate the vast majority of  Johnson-Kron-Cousins observations, but modern CCD detectors and manufactored filters provide slightly different responses. To adhere as much as possible to the Landolt original system, authors in the cited literature applied empirical, {\em ad hoc} non-linear corrections to their measurements. By accurately calibrating our observed magnitudes on the original \citet{LANDOLT1992a} system, we observe no significant trend between our SPSS results and the Landolt or Clem magnitude collections. However, the CALSPEC spectra were calibrated in a completely independent way, without applying any particular empirical re-calibration on the Landolt original reference system. This type of discrepancy between contemporary equipment and the original Landolt one cannot be repaired by a simple wavelength shift in the passbands, at least when the colour range covered is as large as the one of the {\em Gaia} SPSS candidates presented here. It would be necessary to re-compute the photometric passbands so that their shape adequately takes into account the typical deviation of modern CCDs and filter systems from the Landolt one. Among  the passbands provided  in the literature, we found that the \citet{BESSELL2012} ones provide the best results, but they are apparently still not sufficiently accurate for some equipment.

In conclusion, the trends observed between the SPSS and the synthetic magnitudes obtained from CALSPEC spectra were observed in several previous studies. They are not caused by specific problems in the SPSS presented here, but by differences in the CALSPEC equipment with respect to the original Landolt equipment, that have not been explicitly corrected for, unlike in photometric studies. Apart from the trends, the 1--2 per cent offset we found with respect to the CALSPEC synthetic magnitudes is similar in amplitude to the one observed with the Landolt collection, which in turn is larger than the one observed with the Clem collection. The offset goes in the same direction in all comparisons, suggesting that the SPSS  absolutely calibrated magnitudes presented here, while statistically compatible with the literature measurements, are systematically   overestimated by a small quantity that varies in the range 1.0$\pm$0.5 per cent, depending on the comparison sample and on the passband.

\section{Discussion and conclusions}
\label{sec:disc}

We have presented accurately calibrated magnitudes for 228 stars, 198 SPSS candidates and 30 rejected SPSS candidates. Photometry will be used to validate the SPSS spectrophotometry, especially for those SPSS observed in non-optimal conditions. Comparison with literature photometry and spectrophotometry suggests that our magnitudes are systematically underestimated by a small quantity that varies in the range 1.0$\pm$0.5 per cent. The systematic offset is within statistical uncertainties and, what is most important, is comparable to the current, state-of-the-art accuracy as estimated by extensive comparisons among the Landolt, Clem, and CALSPEC collections. However, in the case of space missions like {\em Gaia}, the actual {\em internal} uncertainties on the integrated photometry are much smaller than the {\em external} uncertainty that is achievable using existing systems of standard stars. The internal uncertainties of {\em Gaia} are estimated to be well below the millimagnitude level  in the $ 6 < G < 16$ range and  $\sim 10$ millimag at the bright ($G\simeq 3$) and faint ($G\simeq 20$) ends \citep{Evans18}. So the question is: how can we move towards such small uncertainties in the {\em external} calibration of {\em Gaia} data as well?

Indeed, with current detectors, it is feasible to obtain at least 1 per cent precision in the determination of the physical energy distribution of stars, as well as better than 1 per cent   precision in laboratory reference standard flux measurements \citep{BROWN2006}. Nevertheless, the direct comparison between laboratory irradiance standard sources (usually blackbodies and tungsten lamps of known flux) and standard stars fluxes has not been equally developed. The early works by \cite{OKE1970} and \cite{HAYES1970} and their subsequent revisions and improvements 
(\citealt{HAYES1975,TUG1977,HAYES1985,MEGESSIER1995} and references therein) on measuring the absolute flux of the primary ground-based standard star Vega with respect to  terrestrial standard sources, represent the pillars of this effort. A proper comparison requires the standard source to be far enough from a ground based telescope in order to simulate  a stellar point source with a collimated beam. These measurements are complicated by many effects such as the differential  atmospheric absorption  between the standard lamp and the standard star light (whose correction requires a precise  knowledge of the atmospheric transmission as a function of wavelength), 
the brigthness difference  between the two sources and the need to characterize the telescope/detector throughput. Early results have not been followed by many  modern comparisons of laboratory standards to stars \citep{bohlin14}, even if new methods have been proposed \citep{SMITH2009}, and this technique has been superseded by different approaches, nicely described in \cite{STUBBS2015}, based on calibrated detectors or on theoretical knowledge of the physics of hydrogen. In fact,
as mentioned in \cite{ZIMMER2016}, over the last decades the National Institute for Standards and Technology\footnote{\url{https://www.nist.gov/}} put aside emissive standards in favour of  detector standards, such as silicon and InGaAs photodiodes \citep{LARASON1996,YOON2003}, that have been calibrated against primary optical standards \citep{BROWN2006,SMITH2009}.

On the other hand, the flux distributions of spectrophotometric standard stars are currently based on model atmosphere calculations, as in the HST CALSPEC archive of flux standards, that rely on the calculated model atmospheres of three pure hydrogen white dwarfs star normalized to an absolute flux level based on  visible and IR absolute measurements \citep{BOHLIN2007,bohlin14}. In this case the  uncertainties  on model atmospheres must be accounted for, but there is good evidence that relative fluxes from the visible to the near-IR are currently accurate to $\sim 1$ per cent for the primary reference standards \citep{BOHLIN2020a}. Once the standard stars are established, they can be used for absolute optical calibration under photometric condititions. In this lucky case the issue is no longer if the night is photometric but becomes how  photometric it is, because subtle unnoticed atmospheric variations may affect the observations as well as other issues such as filters or instrumental mismatches. All these small effects can make it very difficult to reach an absolute  calibration accuracy better than the current 1 per cent.

Nowadays detector and instrument capabilities probably have the potential to attain even a sub-percent accuracy, but their full exploitation requires techniques for monitoring the actual throughput of the telescopes \citep{STUBBS2007,REGNAULT2009,DOI2010} and for monitoring the atmospheric layers above the ground-based telescope with a more accurate and complete spatial and temporal sampling than the  classical weather stations. This approach, conceptually similar to the  adaptive optics (AO) philosophy, should be able to provide photometric measurement corrections for direction-, wavelength-, and time-dependent astronomical extinction, as described in 
\cite{MCGRAW2010,ZIMMER2010,ZIMMER2016}, pursuing the development of what we may call "Adaptive Photometry", something similar to AO, but in the photometric regime. In this way, photometric observations can be corrected in real time, taking into account the actual  transmission of the column of air through which calibrations are being made. Many researchers are focusing their efforts in developing methods of real-time direct atmosphere monitoring, but the details are beyond the scope of this paper. Nevertheless we note that the atmospheric calibration system for the Vera C. Rubin Observatory (previously known as LSST) is moving in this direction, including a 1.2\,m dedicated atmospheric calibration telescope and a set of instrumentation to monitor precipitable water vapor and cloud cover \citep{SEBAG2014}. All measurements by the atmospheric calibration telescope will be combined with MODTRAN\footnote{MODerate resolution atmospheric TRANsmission, \url{http://modtran.spectral.com/}} atmospheric models to characterize and monitor the night-time atmospheric properties, while a sun monitoring system will measure the atmospheric aerosols during daytime.

Free from atmospheric disturbances, rockets or ballon experiments  can also boost the accuracy of stars  absolute calibration.
The  Absolute Color Calibration Experiment for Standard Stars (ACCESS) rocket experiments \citep{KAISER2005,KAISER2018} aims at an absolute spectrophotometric 
calibration accuracy of 1 per cent in the 0.35--1.7 $\mu$m range, but due to the short flight time outside 
the atmosphere, observations are limited to a few bight ($\lesssim 10$ mag) stars. High altitude balloons can 
provide long-term astronomical observations with virtually no  atmospheric disturbance, but a suitable 
star tracking and image stabilization may be challenging \citep{FESEN2006}.
Satellites and small stratospheric ballons  can also be used to lift a calibrated source above the atmosphere in order to use it as a standard source to determine the atmospheric transmission, as tested with the Cloud Aerosol Lidar and Infrared Pathfinder Satellite Observations  (CALIPSO, \citealt{WINKER2009})  or as  devised by the Airborne Laser for Telescopic Atmospheric Interference Reduction (ALTAIR) project \citep{ALBERT2016}.

For the moment, however, the 1 per cent accuracy limit appears as the best that can be achieved with current equipment, and the SPSS magnitudes presented here are the best that can be done to assist the external calibration of {\em Gaia}, within the current framework. In the future, when more sophisticated solutions will become available, it will be certainly possible to recalibrate {\em Gaia} data starting from publicly available data.

\section*{Acknowledgements}

We would like to acknowledge the financial support of the Istituto Nazionale di Astrofisica (INAF) and specifically of the Arcetri, Roma, and Bologna Observatories; of ASI (Agenzia Spaziale Italiana) under contract to INAF: ASI 2014-049-R.0 dedicated to SSDC, and under contracts to INAF: ARS/96/77, ARS/98/92, ARS/99/81, I/R/32/00, I/R/117/01, COFIS-OF06-01, ASI I/016/07/0, ASI I/037/08/0, ASI I/058/10/0, ASI 2014-025-R.0, ASI 2014-025-R.1.2015 dedicated to the Italian participation  to the Gaia Data Analysis and 
Processing Consortium (DPAC). 
We acknowledge financial support from the PAPIIT project IN103014 (UNAM, Mexico).
This work was supported by the MINECO (Spanish Ministry of Economy) through grant RTI2018-095076-B-C21 (MINECO/FEDER, UE). 
APV acknowledges FAPESP for the postdoctoral fellowship No. 2017/15893-1 and the DGAPA-PAPIIT grant IG100319.
\\
The results presented here are based on observations 
made with: the ESO New Technology Telescope(NTT) in La Silla, Chile (programmes: 182.D-0287, 086.D-0176,  091.D-0276,  093.D-0197, 094.D-0258);
the Italian Telescopio Nazionale Galileo (TNG) operated on the island of La Palma by the Fundaci\'{o}n Galileo Galilei of INAF (Istituto Nazionale di Astrofisica) at the Spanish `Observatorio del Roque de los Muchachos' of the Instituto de Astrofisica de Canarias (programmes: AOT20\_41, AOT21\_1, AOT29\_Gaia\_013/id7); the Nordic Optical Telescope (NOT), operated by the NOT Scientific Association at the `Observatorio del Roque de los Muchachos', La Palma, Spain, of the Instituto de Astrofisica de Canarias (programmes: 49-013, 50-012); 
the 2.2~m telescope at the Centro Astron\'{o}mico Hispano-Alem\'{a}n (CAHA) at Calar Alto, operated jointly by Junta de Andalucía and Consejo Superior de Investigaciones Científicas (IAA-CSIC); 
the 1.5~m telescope Observatorio Astron\'{o}mico Nacional
at San Pedro M\'{a}rtir, B.C., Mexico; the Cassini Telescope at Loiano, operated by the INAF Astronomical Observatory of Bologna. We take the occasion to warmly thank the staff of these observatories for their support  and we remember in particular  Gabriel Garcia, who died in a fatal accident while working at the Observatorio Astron\'{o}mico Nacional
at San Pedro M\'{a}rtir. \\
We thank the referee  M.S. Bessell for  his helpful comments.
\\
This research has made use of the following
software and online databases: the Aladin Sky Atlas\footnote{\url{https://aladin.u-strasbg.fr/}}
developed at CDS,  Strasbourg Observatory, France \citep{BONNAREL2000, BOCH2014}, CALSPEC, \textsc{iraf}, the NASA's Astrophysics Data System\footnote{\url{https://ui.adsabs.harvard.edu/}},
\textsc{R}\footnote{\url{https://www.r-project.org/}}, \textsc{SAOImage DS9}\footnote{\url{https://sites.google.com/cfa.harvard.edu/saoimageds9}},  
\textsc{SExtractor}\footnote{\url{https://www.astromatic.net/software/sextractor}.}, the Simbad\footnote{\url{http://simbad.u-strasbg.fr/simbad/}} database, operated at CDS, Strasbourg, France \citep{WENGER2000}, \textsc{Supermongo}\footnote{\url{https://www.astro.princeton.edu/~rhl/sm/}.}.

\section{Data availability}
All relevant data  generated  during the current study
and not already available in this article and in its online supplementary material,
will  be made public through the SSDC {\em Gaia} Portal\footnote{\url{http://gaiaportal.ssdc.asi.it/}} with the final release of flux tables, that should correspond to the {\em Gaia} fourth release\footnote{\url{https://www.cosmos.esa.int/web/gaia/release}}. 



\bibliographystyle{mnras}
\bibliography{SPSSGaia13112020} 






\bsp	
\label{lastpage}
\end{document}